\documentclass[aps, prd, onecolumn, amsmath, superscriptaddress, floatfix, nofootinbib, preprintnumbers, preprint]{revtex4-1}

\usepackage{amsfonts}
\usepackage{amsmath, amssymb}
\usepackage{physics}
\usepackage{xcolor}
\usepackage[colorlinks=true,citecolor=blue,linkcolor=blue,urlcolor=blue]{hyperref}
\usepackage{bm}
\usepackage{slashed}
\usepackage{braket}
\usepackage{siunitx}
\usepackage{graphicx}
\usepackage{dcolumn} 

\newcommand{\xb}{x_{\rm B}}
\newcommand{\diff}[1]{\mathrm{d}#1}
\newcommand{\Apv}{A_{\rm PV}}
\newcommand{\Apvd}{\Apv^{\scriptscriptstyle D}}
\newcommand{\ssw}{\sin^2{\theta_{\rm W}}}

\newcommand{\eref}[1]{Eq.~(\ref{eq:#1})}

\newcommand{\fref}[1]{Fig.~\ref{fig:#1}}

\newcommand{\sref}[1]{Sec.~\ref{sec:#1}}
\newcommand{\ssref}[1]{Sec.~\ref{ssec:#1}}

\begin{document}

\title{Impact of parity-violating deep-inelastic scattering \\ 
on the weak mixing angle and high-$x$ parton distributions}

\author{R.~M.~Whitehill}
\affiliation{Department of Physics, Old Dominion University, Norfolk, Virginia 23529, USA}
\affiliation{Jefferson Lab, Newport News, Virginia 23606, USA}
\author{M.~M.~Dalton}
\affiliation{Jefferson Lab, Newport News, Virginia 23606, USA}
\author{T.~Liu}
\affiliation{Key Laboratory of Particle Physics and Particle Irradiation (MOE), Institute of Frontier \\ and Interdisciplinary Science, Shandong University, Qingdao, Shandong 266237, China \\
        \vspace*{0.2cm}
        {\bf JAM Collaboration \\ \footnotesize{~(PDF Analysis Group)}
        \vspace*{0.2cm}}}
\author{W.~Melnitchouk}
\author{N.~Sato}
\affiliation{Jefferson Lab, Newport News, Virginia 23606, USA}

\date{\today}
\preprint{JLAB-THY-26-4810}

\begin{abstract}
We discuss the impact of neutral current parity-violating deep-inelastic scattering (PVDIS) of electrons from protons and deuterons on the determination of the weak mixing angle, $\ssw$, and parton distribution functions (PDFs) at large parton momentum fractions $x$.
Using the JAM global QCD analysis framework, we study the effect of incorporating pseudodata simulated for 11~GeV and 22~GeV Jefferson Lab kinematics, accounting for radiative corrections in a factorized QED+QCD approach and uncertainties from higher twist corrections in $\gamma Z$ exchange.
We find that including future PVDIS pseudodata could yield important constraints on the value of $\ssw$ at low $Q^2$ and on the high-$x$ behavior of the strange quark and $d/u$ PDF ratio.
The strong correlation between $\ssw$ and the $x$ dependence of the PDFs demonstrates the necessity for simultaneous analysis of QCD and electroweak quantities to ensure an unbiased determination of the weak mixing angle from PVDIS data.
\end{abstract}

\maketitle

\section{Introduction}   %
\label{sec:introduction} %

The electroweak sector of the standard model (SM) represents an important background in the quest to understand strongly interacting nuclear phenomena.
The fundamental constants of the SM, such as electroweak couplings, gauge boson masses, and mixing parameters (including the weak mixing angle and the Cabibbo–Kobayashi–Maskawa matrix elements), play an important role in global QCD analysis and influence our understanding of hadron structure.
The weak mixing angle, $\ssw$, in particular, has received considerable attention in the literature over the past two decades, especially with regards to its variation with the energy scale.
The scale dependence arises from the ultraviolet renormalization of the SM Lagrangian, and is governed by a renormalization group equation.
The value of $\ssw$ has been precisely measured at the mass of the $Z$ boson~\cite{ALEPH:2005ab, D0:2011baz, CDF:2014wea, CMS:2011utm, CDF:2018cnj, CMS:2018ktx}, which uniquely determines its running within the SM at all energy scales.

It was rather surprising, therefore, when the extraction of $\ssw$ by the NuTeV Collaboration~\cite{NuTeV:2001whx} from ratios of neutral current to charged current neutrino and antineutrino cross sections from nuclei found a roughly 3 standard deviations discrepancy with the SM value.
This result called into question the validity of the running of $\ssw$ in the SM, and led to speculation about possible new physics beyond the SM that become pronounced at low energy scales.
On the other hand, the analysis of the NuTeV data required various approximations to be made in order to relate the measured neutrino-nucleus cross sections to those for a free nucleon, the veracity of which has been debated by various authors~\cite{Londergan:2003ij, Kulagin:2003wz, Gluck:2005xh, NuTeV:2007uwm, Cloet:2009qs, Bentz:2009yy}, leading to the widely-held view that SM corrections may be responsible for the apparent anomaly.

The latter perspective offers a new avenue for global QCD analyses to extract fundamental theory parameters, such as $\ssw$, and parton distribution functions (PDFs) encoding hadron structure within a more holistic paradigm.
At present, however, there exists only a sparse collection of experimental quantities sensitive to $\ssw$ at low energy scales, as illustrated by the reduced-$\chi^2$, $\chi^2_{\rm red} = \chi^2/N_{\rm dat}$, profile in \fref{rchi2-historical}, leaving $\ssw$ relatively unconstrained at these scales, independent of the high energy extractions, which have a relative uncertainty of $\sim 0.1\%$~\cite{ParticleDataGroup:2020ssz}.
This offers lower energy facilities, such as CEBAF at Jefferson Lab, a unique opportunity to measure observables requiring both electroweak and hadronic input in order to test the validity of the SM at low energies and further our knowledge of hadron structure.

\begin{figure}[t] 
    \centering
    \includegraphics[width=0.7\textwidth]{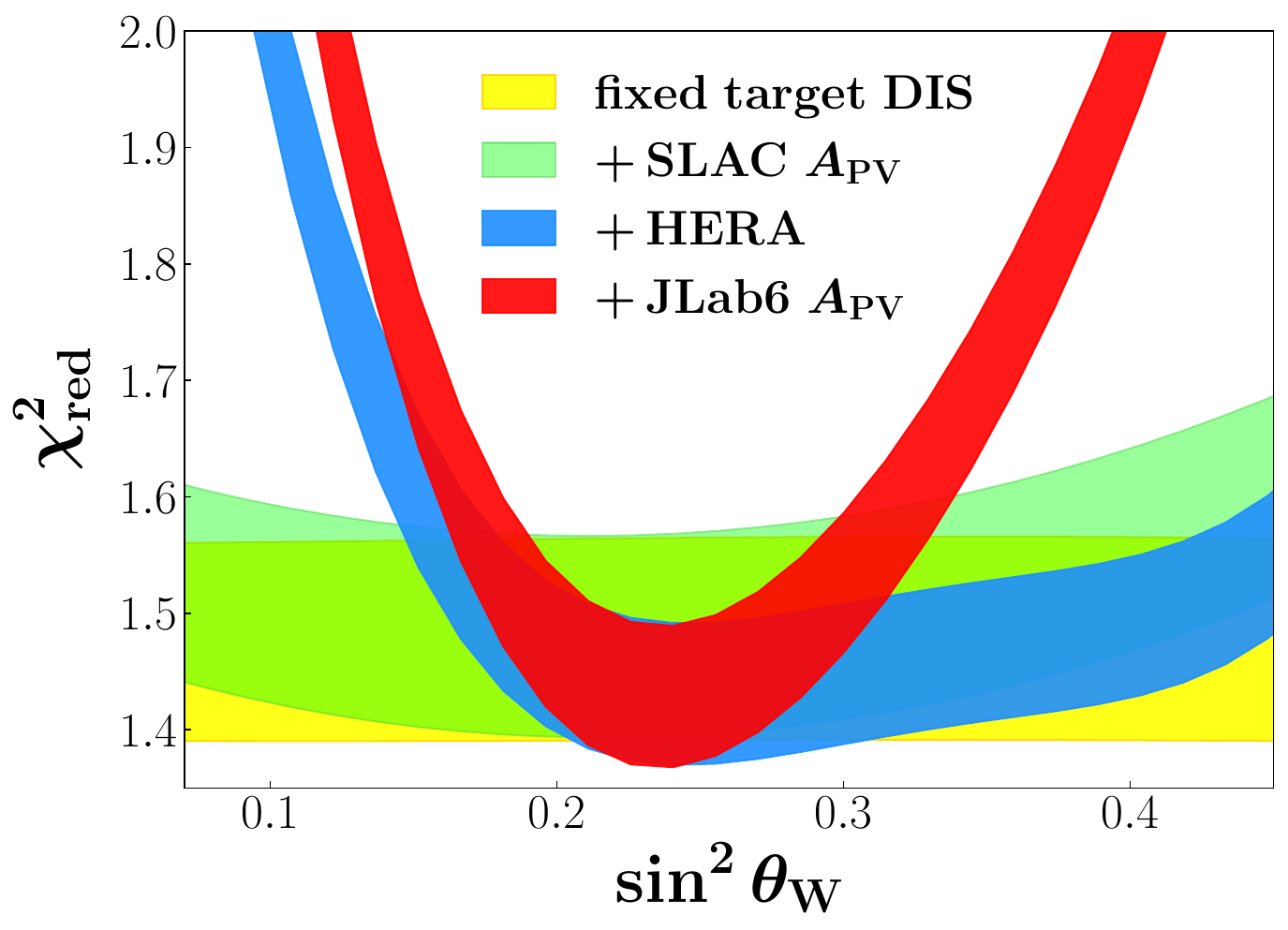}
    \caption{Reduced-$\chi^2$ profiles of $\ssw$ at $\mu^2 = 4$~GeV$^2$ with fixed target DIS datasets (yellow) used in global QCD analyses and previous $A_{\rm PV}$ data from SLAC~\cite{Prescott:1978tm, Prescott:1979dh} (green) and Jefferson Lab~\cite{JeffersonLabSoLID:2022iod} with a 6~GeV beam energy (red), along with HERA data (blue) which include charged and neutral current cross sections. The bands represent $1\sigma$ deviations from the mean.
    }
    \label{fig:rchi2-historical}
\end{figure}

In particular, the SoLID Collaboration at Jefferson Lab has proposed measuring the parity-violating deep-inelastic scattering (PVDIS) asymmetry, $\Apv$, for polarized electrons scattering from protons and deuterons~\cite{JeffersonLabSoLID:2022iod}.
For PVDIS from the deuteron, Bjorken observed~\cite{Bjorken:1978ry} that at leading order (LO) in the strong coupling $\alpha_s$, and assuming dominance of valence quark PDFs at large values of $\xb$, the $\Apv$ asymmetry directly probes the weak mixing angle, $\ssw$.
Indeed, early measurements of $\Apv$ at SLAC~\cite{Prescott:1978tm, Prescott:1979dh} were instrumental in confirming SU(2)$\times$U(1) as the correct theory to describe the electroweak sector of the SM.

With our current understanding of QCD and the strong interactions, more quantitative analysis of PVDIS from the deuteron must include sea quark effects and higher order contributions, which gives rise to additional dependence on the strange and antistrange quark PDFs in $\Apv$.
Understanding the strange sea inside the proton has been an important goal in nuclear physics for many years, and is paramount to advancing our knowledge of nucleon structure~\cite{Kusina:2012vh, Accardi:2016qay, Lai:2007dq, Faura:2020oom, HERMES:2008pug, Jimenez-Delgado:2013sma, Anderson:2024evk}.
Currently, the dependence of the unpolarized strange quark PDF on the parton momentum fraction $x$ remains poorly constrained, with persistent tensions among different global analyses, as illustrated in \fref{s+-dist} for the $s^+ \equiv s + \bar s$ distribution.
Recent data on inclusive $W$ and $W+$charm production in $pp$ collisions from the CMS and ATLAS collaborations at the LHC~\cite{CMS:2016qqr, ATLAS:2023ibp, ATLAS:2012sjl, ATLAS:2016nqi, ATLAS:2014jkm}, along with semi-inclusive hadron production at COMPASS~\cite{COMPASS:2016xvm, COMPASS:2016crr}, have provided evidence of an enhanced strange content in the proton at $x \sim 0.01$, with a suppression at higher $x$ values, $x \sim 0.1$~\cite{Anderson:2024evk}.
On the other hand, of notable importance is the possibility of a strange sea asymmetry, $s^- \equiv s - \bar s$, which has been suggested by both theoretical models and neutrino DIS and dimuon data~\cite{Catani:2004nc, Mason:2006qa, NuTeV:2001dfo, Olness:2003wz, Brodsky:1996hc, Sufian:2018cpj,Li:2001nv, Alwall:2004rd, Melnitchouk:1999mv, Signal:1987gz, PhysRevD.45.958}.
Although some global analyses find deviations of $s^-$ away from zero, no conclusive evidence yet confirms this~\cite{Anderson:2024evk, Hou:2019efy, NNPDF:2021njg, Bailey:2020ooq}.
While the small-$x$ region and issues related to this sea asymmetry remain a significant challenge, likely requiring higher-energy experiments or other nonperturbative input~\cite{Yang:2018nqn, Alexandrou:2020sml}, the intermediate-$x$ and valence regions are more readily accessible through global analysis of experimental data from existing, planned, and proposed programs at facilities such as Jefferson Lab and the Electron-Ion Collider.

\begin{figure}[t] 
    \centering
    \includegraphics[width=\textwidth]{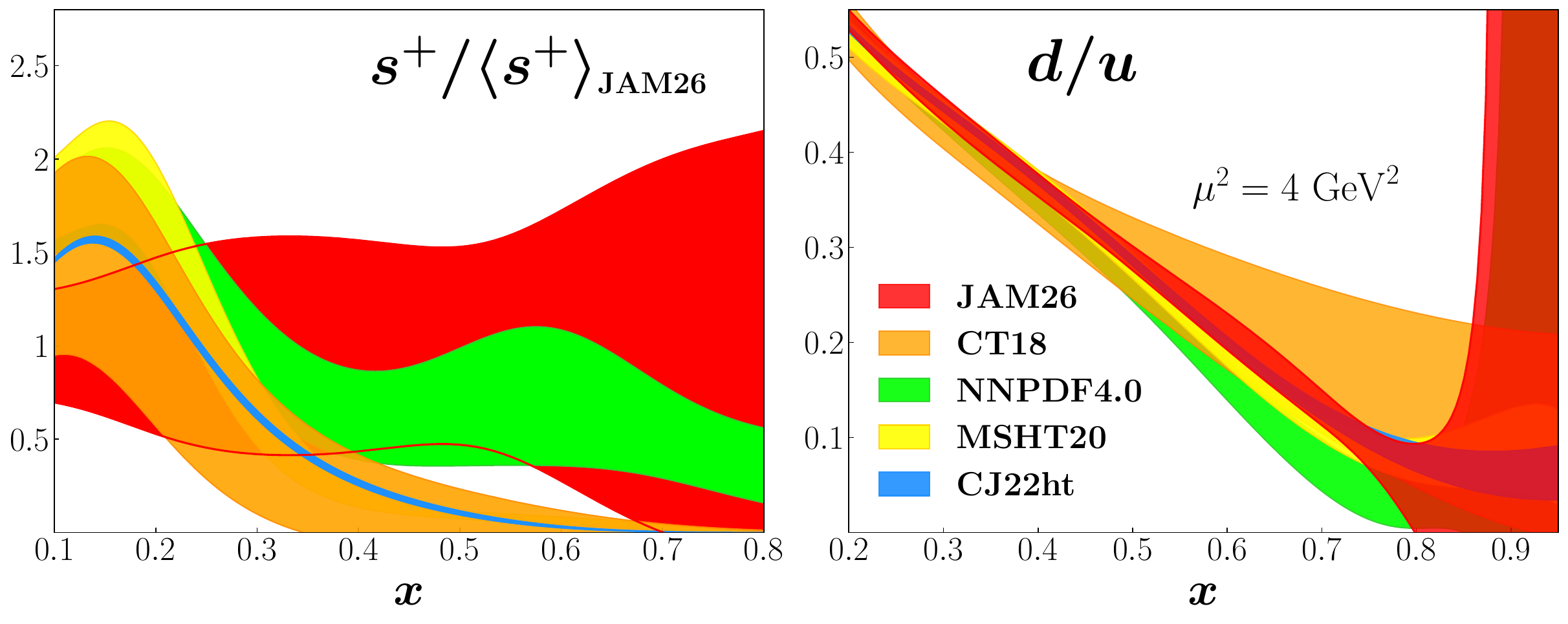}
    \caption{Strange quark PDF $s^+$ relative to the mean JAM26~\cite{Cocuzza:2026zoy} value (left) and $d/u$ ratio (right) versus parton momentum fraction $x$ at $\mu^2 = 4$~GeV$^2$. The JAM26 results (red bands) are compared with the CT18~\cite{Hou:2019efy} (orange), NNPDF4.0~\cite{NNPDF:2021njg} (green), MSHT20~\cite{Bailey:2020ooq} (yellow), and CJ22ht~\cite{Accardi:2023gyr} (blue) PDF parametrizations. The JAM26, NNPDF4.0, and CJ22ht bands represent $1\sigma$ deviations from the average over replicas, while the CT18 band is computed using the Hessian formalism.}
    \label{fig:s+-dist}
\end{figure}

Another important challenge in PDF phenomenology is that of the ratio of down-quark to up-quark PDFs in the limit where the parton momentum fraction $x \to 1$, which provides a window on the nonperturbative dynamics of valence quarks and the role of diquark correlations in the proton~\cite{Melnitchouk:1995fc}.
Unfortunately, the kinematic reach of existing DIS data extends only to $\xb \lesssim 0.8$, so extrapolating $d/u$ to $x \approx 1$ is therefore strongly sensitive to model assumptions and parameterization bias, preventing any definitive conclusion to be made about the $x \to 1$ behavior~\cite{JeffersonLabHallATritium:2021usd, CLAS:2014jvt, Li:2023yda, Cocuzza:2021rfn, Hague:2023tqs, Cocuzza:2026vey, Cocuzza:2026zoy}, as illustrated in \fref{s+-dist}.
Data from $\nu$-DIS and charged current DIS processes also probe $d/u$ through flavor mixing in the weak charged current, but limited statistical precision and uncertainties from nuclear effects and electroweak inputs weaken and contaminate the desired signal~\cite{Radescu:2004zn, NuTeV:1998wnx, NuTeV:1999rwn, NuTeV:2002doy, MINERvA:2014rdw, Cruz-Martinez:2023sdv, DUNE:2021tad}.
Below, we discuss sensitivity of the parity-violating asymmetry on a proton target to the $d/u$ ratio, thus potentially providing a cleaner signal from which to determine $d/u$ at large values of the parton momentum fractions.

In this paper, we explore the sensitivity of the $\Apv$ observables to $\ssw$, the strange quark PDF, and the $d/u$ ratio, and carry out global analyses at next-to-leading order (NLO) in $\alpha_s$ within the JAM Bayesian MC analysis framework.
In particular, we include for the first time the weak mixing angle in the simultaneous fitting paradigm, with $\Apv$ pseudodata simulated for the existing 11~GeV Jefferson Lab facility, as well as a possible future 22~GeV upgrade.
We begin our presentation in \sref{theoretical-framework} with a summary of the theoretical framework used to compute the neutral current inclusive DIS cross sections and $\Apv$.
The sensitivity of $\Apv$ to the strange PDFs, down-to-up quark PDF ratio, and weak mixing angle is discussed in \ssref{sensitivity-on-strange-PDFs-ssw}, while in \ssref{QED+QCD-factorization} we outline radiative corrections to the DIS cross sections presented in \ssref{physical-observables}.
In \sref{global-analysis-methodology} we review the fitting methodology used in our global analysis, based on the Bayesian MC global analysis framework.
In \sref{phenomenology} we focus on the generation of pesudodata for our global analyses and the impact of these fits on the Weinberg angle and the strange PDFs.
Finally, we draw conclusions about our results in \sref{conclusion}, and outline some extensions of this work and future directions.

\section{Theoretical framework}   %
\label{sec:theoretical-framework} %

In this section we present the theoretical framework on which this analysis is based.
We begin by summarizing the physical observables relevant for PVDIS, and identifying their sensitivity to the Weinberg angle, and to the strange quark PDF and $d/u$ quark PDF ratio. 
We also review the basic elements of the factorized QED+QCD formalism introduced recently to self-consistently extract partonic correlation functions in the presence of electroweak radiation, and discuss the implementation of higher twist (HT) corrections, the sensitivity to which we explore at the 11~GeV and 22~GeV energies.

\subsection{Physical observables} %
\label{ssec:physical-observables} %

In this work we focus on the inclusive neutral current deep-inelastic scattering (DIS) of a charged lepton (four-momentum $\ell$) from a nucleon target (four-momentum~$P$) to a final state lepton (four-momentum~$\ell'$) and unobserved hadrons $X$ ($p_X$), $e^{\pm}(\ell) + N(P) \to e^{\pm}(\ell') + X(p_X)$.
The invariant center of mass energy squared of the system is $s = (P + \ell)^2 = M^2 + Q^2/(\xb\, y)$, where $M$ is the nucleon mass, $Q^2 = -q^2 = -(\ell - \ell')^2$ is the four-momentum transfer squared of the exchanged boson, $\xb = Q^2/(2 P \cdot q)$ is the Bjorken scaling variable, and $y = (P \cdot q)/(P \cdot \ell)$ is the lepton inelasticity.
The invariant mass squared of the unobserved hadronic final state is $W^2 \equiv p_X^2 = (P + q)^2 = M^2 + Q^2 (1/\xb - 1)$.

In the one-boson exchange approximation, the cross section is given by a tensor product of lepton and hadron tensors,
\begin{eqnarray}
\label{eq:xsec-xy}
    \frac{\diff \sigma_{\lambda_{\ell} S}}{\diff \xb \, \diff y} 
    = \frac{2 \pi \alpha^2 y}{Q^4} \sum_{i} \eta_{i}\, L_{\mu\nu}^{i} W_{i}^{\mu\nu}
,\end{eqnarray}
where $\alpha$ is the electromagnetic fine structure constant, and the sum over $i \in \{ \gamma, \gamma Z, Z \}$ includes the pure-photon exchange, $\gamma Z$ interference, and pure-$Z$ exchange contributions.
The helicity of the incoming lepton is denoted by $\lambda_\ell$, and $S$ represents the spin polarization of the nucleon (in practice a proton or deuteron).
The kinematic factors,
\begin{align}
\label{eq:eta-factors}
    \eta_{\gamma} = 1, 
    \qquad 
    \eta_{\gamma _Z} = \frac{M_Z^2 G_F}{2 \sqrt{2} \pi \alpha}\frac{Q^2}{Q^2 + M_Z^2}, 
    \qquad 
    \eta_{Z} = \eta_{\gamma Z}^2,
\end{align}
collect vertices and propagator factors, where $M_Z$ is the $Z$-boson mass and $G_F$ is the Fermi constant.
The lepton tensors for photon-exchange, $\gamma Z$ interference, and $Z$-exchange are (for massless leptons) given by
\begin{subequations}
\begin{align}
\label{eq:leptonic-tensor}
    L_{\mu\nu}^{\gamma} &= 2\, \big( \ell_\mu \ell_\nu' + \ell_\mu' \ell_\nu - g_{\mu\nu}\, \ell \cdot \ell' - i \lambda_\ell\, \epsilon_{\mu\nu\alpha\beta}\, \ell^{\alpha} \ell'^{\beta} \big), 
    \\
    L_{\mu\nu}^{\gamma Z} &= -(g_V^e - \mathcal{N}_\ell\, \lambda_\ell\, g_A^e)\, L_{\mu\nu}^{\gamma}, 
    \\
    L_{\mu\nu}^{Z} &= \phantom{-}(g_V^e - \mathcal{N}_\ell\, \lambda_\ell\, g_A^e)^2\, L_{\mu\nu}^{\gamma}
,\end{align}
\end{subequations}
where the coefficients collect the electroweak couplings from each contribution relative to the pure-photon exchange channel, with the weak axial-vector and vector couplings to the electron given by $g_A^e = -1/2$ and  $g_V^e = -1/2 + 2 \ssw$, respectively, and $\mathcal{N}_{\ell} = +1 (-1)$ for leptons (antileptons).
Since the hadronic current matrix elements are not exactly known from QCD, we parameterize the hadronic tensor for each channel $i \in \{ \gamma, \gamma Z, Z \}$ in terms of unpolarized structure functions $F_k^i(x,Q^2)$  as~\cite{ParticleDataGroup:2020ssz},
\begin{eqnarray}
\label{eq:hadronic-tensor-structure-functions}
\begin{aligned}
    W^{\mu\nu}_i
    &= -\widetilde{g}^{\mu\nu} F_1^i(\xb,Q^2) 
    + \frac{\widetilde{P}^{\mu}\widetilde{P}^{\nu}}{P \cdot q} F_2^i(\xb,Q^2) 
    + i \epsilon^{\mu\nu\alpha\beta} \frac{P_{\alpha} q_{\beta}}{2 P \cdot q} F_3^i(\xb,Q^2),
\end{aligned}
\end{eqnarray}
where $\displaystyle \widetilde{g}^{\mu\nu} = g^{\mu\nu} - q^{\mu}q^{\nu}/{q^2}$ and $\widetilde{P}^{\mu} = \widetilde{g}^{\mu\nu} P_{\nu}$.

Experimentally, PVDIS typically measures the parity-violating asymmetry, $\Apv$, defined as the ratio between the lepton-polarized and unpolarized cross sections,
\begin{align}
\label{eq:Apv-definition}
    \Apv = \frac{\diff \sigma_{LU}}{\diff \sigma_{UU}}.
\end{align}
Summing over all the possible leptonic and target longitudinal polarization combinations, the unpolarized cross section can be written as
\begin{align}
\label{eq:sigUU}
    \frac{\diff \sigma_{UU}}{\diff \xb \, \diff y} 
    &= \frac{1}{4}\bigg[ \frac{\diff \sigma_{++}}{\diff \xb \, \diff y} + \frac{\diff \sigma_{+-}}{\diff \xb \, \diff y} + \frac{\diff \sigma_{-+}}{\diff \xb \, \diff y} + \frac{\diff \sigma_{--}}{\diff \xb \, \diff y} \bigg] 
    \nonumber \\
    &= \frac{4 \pi \alpha^2}{\xb\, y\, Q^2} 
    \bigg[ K_1(\xb,y,Q^2)\, F_1^{UU}(\xb,Q^2) 
         + K_2(\xb,y,Q^2)\, F_2^{UU}(\xb,Q^2) 
    \nonumber \\
    &\hspace{2cm} 
         + \mathcal{N}_{\ell}\, K_3(\xb,y,Q^2)\, F_3^{UU}(\xb,Q^2) \bigg]
,\end{align}
where the kinematic coefficients are 
\begin{subequations}
\begin{eqnarray}
    K_1(\xb,y,Q^2) &=& \xb\, y^2,   \\
    K_2(\xb,y,Q^2) &=& 1 - y - \frac{\xb^2\, y^2 M^2}{Q^2}, \\
    K_3(\xb,y,Q^2) &=& \xb\, y\, \big( 1 - \tfrac12 y \big),
\end{eqnarray}
\end{subequations}
and unpolarized structure functions given by
\begin{subequations}
\begin{align}
    F_{1}^{UU} &= F_1^{\gamma} - g_V^e \eta_{\gamma Z} F_1^{\gamma Z} + \big[ (g_V^e)^2 + (g_A^e)^2 \big] \eta_Z F_1^Z, \\
    F_{2}^{UU} &= F_2^{\gamma} - g_V^e \eta_{\gamma Z} F_2^{\gamma Z} + \big[ (g_V^e)^2 + (g_A^e)^2 \big] \eta_{Z} F_2^Z, \\
    F_{3}^{UU} &= \phantom{F_3^\gamma} - g_A^e \eta_{\gamma Z} F_3^{\gamma Z} + 2 g_V^e g_A^e \eta_Z F_3^Z.
\end{align}
\end{subequations}
Similarly, the lepton-polarized cross section can be written as
\begin{align}
\label{eq:sigLU}
    \frac{\diff \sigma_{LU}}{\diff \xb \, \diff y} 
    &= \frac14 \bigg[ \bigg( \frac{\diff \sigma_{++}}{\diff \xb \, \diff y} + \frac{\diff \sigma_{+-}}{\diff \xb \, \diff y} \Bigg) - \Bigg( \frac{\diff \sigma_{-+}}{\diff \xb \, \diff y} + \frac{\diff \sigma_{--}}{\diff \xb \, \diff y} \bigg) \bigg] 
    \nonumber \\
    &= -\frac{4 \pi \alpha^2}{\xb\, y\, Q^2} 
    \bigg[ \mathcal{N}_{\ell}\, K_1(\xb,y,Q^2)\, F_1^{LU}(\xb,Q^2) 
         + \mathcal{N}_{\ell}\, K_2(\xb,y,Q^2)\, F_2^{LU}(\xb,Q^2) 
    \nonumber \\
    &\hspace{2.5cm} 
         + K_3(\xb,y,Q^2)\, F_3^{LU}(\xb,Q^2) 
    \bigg]
,\end{align}
where the lepton-polarized structure functions are given by
\begin{subequations}
\begin{align}
    F_{1}^{LU} &= - g_A^e \eta_{\gamma Z} F_1^{\gamma Z} + 2 g_V^e g_A^e \eta_{Z} F_1^Z,\\
    F_{2}^{LU} &= - g_A^e \eta_{\gamma Z} F_2^{\gamma Z} + 2 g_V^e g_A^e \eta_Z F_2^Z, \\
    F_{3}^{LU} &= - g_V^e \eta_{\gamma Z} F_3^{\gamma Z} + \big[ (g_V^e)^2 + (g_A^e)^2 \big] \eta_{Z} F_3^Z.
\end{align}
\end{subequations}
At low $Q^2$ values, $Q^2 \ll M_Z^2$, the $\eta_Z$ term can usually be neglected.

\subsection{Sensitivity to nucleon PDFs and \boldmath{$\ssw$}} %
\label{ssec:sensitivity-on-strange-PDFs-ssw}                    %

In this section, we motivate the study of $\Apv$ at Jefferson Lab in relation to the weak mixing angle, nucleon strangeness, and the $d/u$ ratio by studying the form of \eref{Apv-definition} under some approximations.
First, we write the LO term in the expansion of $\Apv$ in $G_F/\alpha$, specializing to the case of an electron beam such that $\mathcal{N}_{\ell} = 1$.
This effectively leaves us with the interference term in the numerator from $\diff \sigma_{LU}$ and the pure photon exchange term in the denominator from $\diff \sigma_{UU}$, giving
\begin{align}
\label{eq:LO-Apv}
    \Apv = \frac{G_F Q^2}{4 \sqrt{2} \pi \alpha} \Big[ a_1(\xb,Q^2) Y_1(\xb,y,Q^2) + a_3(\xb,Q^2) Y_3(\xb,y,Q^2) \Big]
,\end{align}
where
\begin{subequations}
\begin{align}
    a_1 = 2 g_A^e \frac{F_1^{\gamma Z}}{F_1^\gamma} &, 
    \quad 
    Y_1 = \Bigg( \frac{1 + R^{\gamma Z}}{1 + R^\gamma} \Bigg) \frac{1 + (1 - y)^2 - \frac12 y^2 \Big[ 1 + r^2 - 2r^2/\big(1 + R^{\gamma Z}\big) \Big]}{1 + (1 - y)^2 - \frac12 y^2 \Big[ 1 + r^2 - 2r^2/\big(1 + R^{\gamma}\big) \Big]},
\\
    a_3 = \phantom{2} g_V^e \frac{F_3^{\gamma Z}}{F_1^{\gamma}} &,
    \quad 
    Y_3 = \Bigg( \frac{1 + R^{\gamma Z}}{1 + R^\gamma} \Bigg) \frac{1 - (1 - y)^2}{1 + (1 - y)^2 - \frac12 y^2 \Big[ 1 + r^2 - 2r^2/\big(1 + R^{\gamma}\big) \Big]},
\end{align}
\end{subequations}
where $r^2 = 1 + 4 M^2 \xb^2 / Q^2$ and
\begin{align}
    R^{i} = \frac{F_2^{i}}{2 \xb F_1^i}\, r^2 - 1
\end{align}
is the ratio of the longitudinal to transverse photoabsorption cross sections.

In the leading twist (LT) approximation, the structure functions can be written at LO (in which $x = \xb$) as
\begin{subequations}
\begin{align}
\label{eq:unpolarized-LO-structure-functions}
    F_{1}^{[\gamma,\gamma Z,Z]}(\xb,Q^2) 
    \ &\approx\ \frac12 \sum_q 
    \Big[ e_q^2, \, 2e_{q}g_{V}^{q}, \, (g_{V}^{q})^2+(g_{A}^{q})^2 \Big]\,
    q^+(\xb,Q^2), \\
    F_{3}^{[\gamma,\gamma Z,Z]}(\xb,Q^2) 
    \ &\approx\ \sum_q \Big[ 0, \, 2e_q g_A^q, \, 2g_V^q g_A^{q} \Big]\,
    q^-(\xb,Q^2)
,\end{align}
\end{subequations}
where $q^\pm = q\, \pm\, \bar q$ is the $C$-even or $C$-odd quark flavor combination, $e_q$ is the quark charge, $g_A^q = \frac12 {\rm sgn}(e_q)$ is the axial-vector coupling for a quark of flavor $q$, and $g_V^q = \frac12 {\rm sgn}(e_q) - 2 e_q \ssw$ is the corresponding vector coupling.
At LO the $F_1^i$ and $F_2^i$ structure functions are related by the Callan-Gross relation, $F_2^i = 2 \xb F_1^i$.
At large values of $Q^2 \gg M^2$, one also has 
    $r^2 \approx 1$, 
    $Y_1 \approx 1$, and 
    $Y_3 \approx \big[ 1 - (1 - y)^2 \big]/\big[ 1 + (1 - y)^2 \big]$,
which also gives
    $R^i \approx 0$.
The parity-violating asymmetry then simplifies to
\begin{align}
\label{eq:Apv-LO-alphaS}
    \Apv \approx \frac{G_F Q^2}{2 \sqrt{2} \pi \alpha} 
    \frac{\sum_q e_q \big[ 2 g_A^e g_V^q\, q^+ + 2 g_V^e g_A^q\, Y_3\, q^- \big]}{\sum_q e_q^2\, q^+}
.\end{align}

For the case of an isoscalar deuteron target, neglecting sea quark asymmetries ($s^- = c^- = 0$), the deuteron parity-violating asymmetry can be written as~\cite{Wang:2014guo}
\begin{align}
\label{eq:Apv-D-LO}
    \Apvd &\approx \frac{3 G_F Q^2}{2 \sqrt{2} \pi \alpha} \bigg[ \frac{ 2 g_A^e \big[ (2 g_V^u - g_V^d) + 2 g_V^u R_c - g_V^d R_s \big] + 2 g_V^e (2 g_A^u - g_A^d) Y_3 R_V}{5 + R_s + 4 R_c} \bigg]
,\end{align}
where 
    $R_s = 2 s^+/(u^+ + d^+)$,
    $R_c = 2 c^+ / (u^+ + d^+)$ and 
    $R_V = (u^- + d^-)/(u^+ + d^+)$.
Further expanding \eref{Apv-D-LO} to linear powers of $R_s$, and using the fact that $R_c \ll R_s$ and $Y_3\, R_V \ll R_s$ across the kinematics of interest at Jefferson Lab, we obtain
\begin{align}
\label{eq:Apv-no-PDFs}
    \Apvd \approx -\frac{G_F Q^2}{4 \sqrt{2} \pi \alpha} 
            \bigg[ \bigg( \frac{9}{5} - 4 \ssw \bigg) + \frac{R_s}{25} \bigg].
\end{align}
This makes clear that the dominant term in $\Apvd$ depends on $\ssw$~\cite{Bjorken:1978ry}, while an additional contribution comes from the strange quarks.

In \fref{Apv-strange-uncertainty} the deuteron asymmetry $\Apvd/Q^2$ is shown versus $\xb$ at $Q^2=4$~GeV$^2$ for an incident electron energy $E=11$~GeV for the full NLO calculation.
For reference, we also show the naive parton model result, obtained in the LO approximation in \eref{Apv-no-PDFs}, as obtained in Ref.~\cite{Bjorken:1978ry}.
Although kinematically suppressed, the $a_3 \, Y_3$ term shifts the value of $\Apv^{D}$ and introduces a nontrivial shape in $\xb$, while NLO corrections also induce non-negligible but relatively smaller shifts.
This demonstrates that PDFs play an important role in the description of $\Apv$ and must be included in any quantitative analysis of future deuteron data.
The full NLO result is also compared with the asymmetry obtained when setting the strange quark PDF to zero, which yields an artificially smaller uncertainty band due to the large $s^+$ PDF uncertainty in the full result.
While the central value of the predicted $\Apvd$ is not very sensitive to changes in $s^+$, the theoretical uncertainties on the predicted $\Apvd$ are dominated by the strange PDF uncertainties.
Future $\Apvd$ data could therefore serve as a unique and relatively clean observable for inclusion in global QCD analyses to better constrain the $s^+$ PDF.

\begin{figure}[t] 
    \centering
    \includegraphics[width=0.7\textwidth]{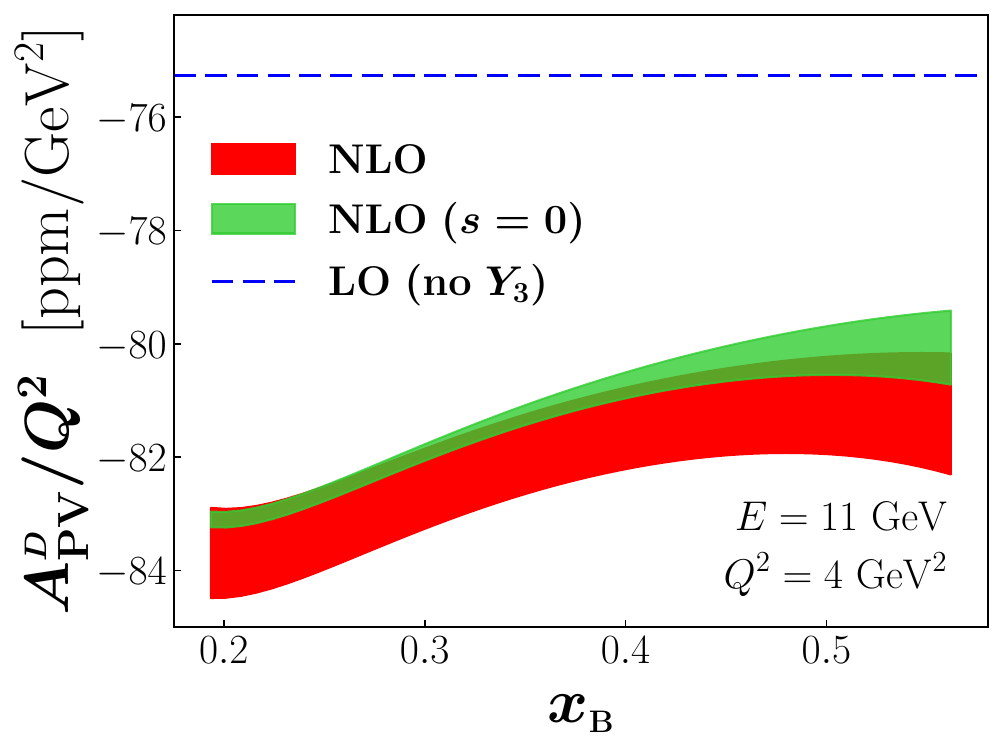}
    \caption{Predictions for the deuteron parity-violating asymmetry $\Apvd$ with an electron beam energy of 11~GeV and $Q^2 = 4$~GeV$^2$ at NLO using JAM PDFs with (red band) and without (green band) the strange channel, and compared also with the LO result with setting the $Y_3$ term in \eref{Apv-D-LO} to zero (blue dashed line).}
    \label{fig:Apv-strange-uncertainty}
\end{figure}

For the case of a proton, neglecting the kinematically suppressed $Y_3$ term, and keeping only the $u$ and $d$ flavor contributions in both the numerator and denominator, we obtain
\begin{align}
\label{eq:LO-Apv-proton-d/u}
    \Apv^{p} \approx -\bigg( \frac{3 G_F Q^2}{4 \sqrt{2} \pi \alpha} \bigg) \, \frac{g_V^{u} - \tfrac12 g_V^d \, (d/u) }{1 + \tfrac14 \, (d/u)}
,\end{align}
which shows the dependence of the proton PVDIS asymmetry on the $d/u$ PDF ratio.
In traditional inclusive DIS measurements, $d/u$ can be determined from the ratio of the neutron and proton $F_2^{\gamma}$ structure functions.
However, since stable, free neutron targets or beams do not exist, alternative nuclear targets, such as deuterium, must be used instead, thus entangling $d/u$ with nuclear effects in global analyses and rendering phenomenological extractions more challenging.
On the other hand, \eref{LO-Apv-proton-d/u} demonstrates that direct sensitivity to $d/u$ exists in the proton PVDIS asymmetry, independent of nuclear effects, which could provide valuable input into future global analyses.

\subsection{QED+QCD factorization} %
\label{ssec:QED+QCD-factorization} %

In reactions involving large momentum transfer, photon radiation is triggered by the incident and scattered leptons and quarks, which can significantly alter the description of the DIS process.
Radiative effects and their associated corrections have been extensively studied in the literature, and most existing treatments modify the cross section multiplicatively with kinematic smearing, use Monte Carlo event generators to remove radiative effects, or distort the contributions from radiative effects and hadron structure~\cite{Mo:1968cg, Bardin:1976qa, Badelek:1994uq}.
In practice, collinear radiative emissions from the incoming and outgoing leptons can lead to substantial modifications of the dynamics of the scattering process.

Formally, the QED radiative effects can be treated perturbatively, jointly alongside the QCD interactions, which leads to a tower of corrections of $\mathcal{O}(\alpha^{m} \alpha_s^{n})$ beyond the typical Born-level amplitude, where only a single photon is exchanged.
A complication of this approach, however, is that the photons can also be radiated and absorbed by the very hadrons whose structure we are attempting to probe, and can couple to $q\bar q$ states in addition to $\ell\bar\ell$ pairs, which introduces additional nonperturbative elements into the analysis.
To avoid possible inconsistencies between hadronic structure inputs and outputs, Liu {\it et al.}~\cite{Liu:2020rvc, Liu:2021jfp} introduced a hybrid QED+QCD factorization formalism in which the partonic distributions in DIS are extracted at the same time as the leptonic distributions, with QED and QCD effects accounted for simultaneously.
In this case, the differential cross section for the DIS process can be written as
\begin{align}
    \frac{E_{\ell'}\, \diff\sigma_{\ell P \to \ell' X}}{\diff^3 \ell'} 
    = \int_{\zeta_{\rm min}}^{1} \frac{\diff \zeta}{\zeta^2}\,
    D_{e/e}(\zeta,\mu^2) \int_{\xi_{\rm min}}^{1} \diff \xi \, f_{e/e}(\xi,\mu^2)\, 
    \frac{E_{k'} \diff\sigma_{k P \to k' X}}{\diff^3 k'} \Bigg|_{k=\xi \ell, k'=\ell'/\zeta}
,\end{align}
where $f_{e/e}(\xi,\mu^2)$ and $D_{e/e}(\zeta,\mu^2)$ are the lepton distribution function (LDF) and lepton fragmentation function (LFF), respectively, which can be interpreted as probabilities to find an electron with momentum fraction $\xi$ or $\zeta$, respectively, in the incident or fragmenting electron, with minimum values
\begin{subequations}
\begin{align}
    \xi_{\rm min} &= \frac{1-y}{\zeta - \xb\, y},
    \\
    \zeta_{\rm min} &= 1- (1- \xb)\, y
.\end{align}
\end{subequations}
Transforming to a cross section differential in $\xb$ and $y$, we can write the differential cross sections in Eqs.~(\ref{eq:sigUU}) and (\ref{eq:sigLU}) as
\begin{align}
\label{eq:QED-xsec-xy}
    \frac{\diff \sigma_{(U/L)U}}{\diff \xb \, \diff y} = \int_{\zeta_{\rm min}}^{1} \frac{\diff \zeta}{\zeta^2} D_{e/e}(\zeta,\mu^2) \int_{\xi_{\rm min}}^{1} \diff \xi \, f_{e/e}(\xi,\mu^2) 
    \bigg[ \frac{Q^2}{\xb} \frac{\hat{x}_{\rm B}}{\hat{Q}^2} \bigg] 
    \frac{\diff \hat{\sigma}_{(U/L)U}}{\diff \hat{x}_{\rm B} \, \diff \hat{y}}\, ,
\end{align}
where $\diff \hat{\sigma}_{(U/L)U} = \diff \sigma_{(U/L)U}(k=\xi \ell,k' = \ell'/\zeta)$, and the hatted kinematic variables are defined in analogy to the hadronic DIS quantities defined above.

Importantly, the operator structure of the LDFs and LFFs is very similar to that of the PDFs and parton to hadron FFs, respectively, so that properties such as their scale evolution follow immediately from the DGLAP evolution equations.
Neglecting hadronic contributions, these functions can be also computed perturbatively in QED.
In this work we utilize the LO LDFs and LFFs, which are given at the input scale $\mu_0$ by
\begin{subequations}
\begin{align}
    f_{e/e}(\xi,\mu_0^2)   &= \delta(1 - \xi), \\
    D_{e/e}(\zeta,\mu_0^2) &= \delta(1 - \zeta)
.\end{align}
\end{subequations}

The effect of the QED corrections to the parity-violating asymmetry can be seen in \fref{Apv-RC}.
Generally, as the inelasticity $y$ becomes larger, the radiative phase space, governed by $(\xi,\zeta)$, becomes larger and therefore also the cross sections.
The unpolarized cross section $\diff \sigma_{UU}$ grows much faster in magnitude than the polarized cross section $\diff \sigma_{LU}$ as $y \to 1$ when the radiative corrections are accounted for, so that the asymmetry $\Apv \to 0$ in this limit.
Increasing the electron beam energy would therefore permit access to a larger kinematic phase space where the radiative effects are smaller, and approximately preserve the typical description of the parity-violating asymmetry.
At such kinematics, the LDFs and LFFs do not play as dominant a role in the description of $\Apv$.
As illustrated in \fref{Apv-RC}, higher beam energies provide the largest range of $\xb$ where the inelasticity remains small and $W^2$ is large enough for a factorized description of the process, enlarging the kinematic range over which uncertainties from radiative effects are minimized.

\begin{figure}[t] 
    \centering
    \hspace*{-0.4cm}
    \includegraphics[width=1.02\textwidth]{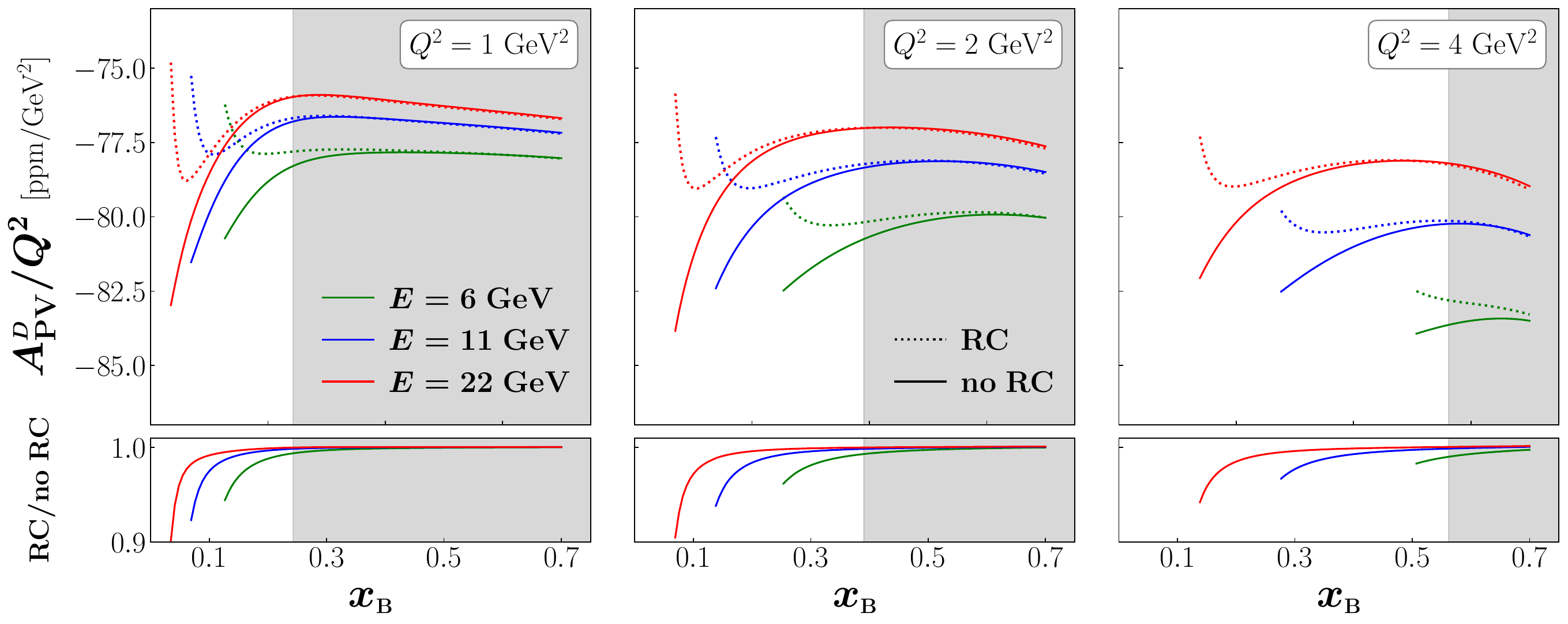}
    \caption{Deuteron PVDIS asymmetry $\Apvd$ versus $\xb$ at several $Q^2$ and beam energy $E$ values, with (dashed lines) and without (solid lines) radiative corrections (RC), along with the ratio (bottom panels) of the corrected to uncorrected values. The gray region spans values of $\xb$ for which $W^2 < 4$~GeV$^2$ that would be excluded from the global analysis. The maximum $\xb$ values shown correspond to $y = 0.7$.}
    \label{fig:Apv-RC}
\end{figure}

\subsection{Higher twist effects}      %
\label{ssec:higher-twist-effects}      %

At the center of mass energies considered in this work, the errors of order $\mathcal{O}\big(\xb^2 M^2/Q^2\big)$ from the leading twist factorization can become non-negligible in regions of the fixed target DIS phase space where $Q^2$ becomes comparable to the intrinsic mass scales, $Q^2 \sim M^2$, or when $\xb \sim 1$.
There are several effects that can contribute to these errors, leading to a tower of power corrections that can modify the leading twist results.
The most studied of these are finite mass effects, such as target mass corrections (TMCs) or heavy quark effects, the origin and phenomenological implications of which have previously been discussed in the literature~\cite{Greenberg:1971lpf, Nachtmann:1973mr, Georgi:1976ve, Ellis:1982cd, Aivazis:1993kh, Moffat:2019qll, Schienbein:2007gr, Accardi:2008ne}.
For consistency with global QCD analysis of various high energy scattering processes, we implement TMCs using the collinear factorization framework~\cite{Aivazis:1993kh, Moffat:2019qll}, which exactly relates the structure function dependence on $\xb$ to that of the Nachtmann scaling variable~\cite{Greenberg:1971lpf, Nachtmann:1973mr}, without making assumptions about the nonperturbative dynamics of the target.
Similarly, heavy quark masses can be treated in the ACOT scheme~\cite{Aivazis:1993kh, Aivazis:1993pi}, although these are less important at the kinematics of interest for this analysis.

\begin{figure}[t] 
    \centering
    \includegraphics[width=\textwidth]{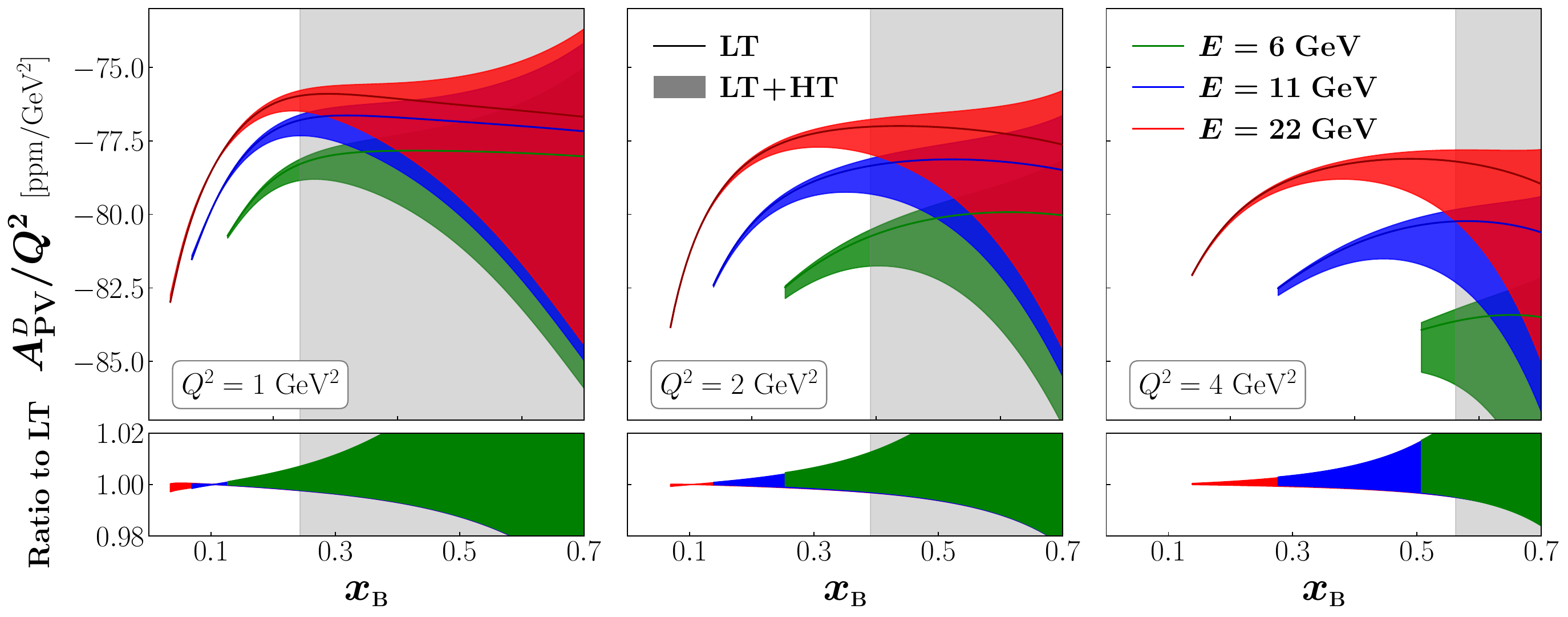}
    \caption{Parity-violating deuteron asymmetry $\Apvd$ as a function of $\xb$ at $Q^2=1$, 2 and 4~GeV$^2$ and beam energies $E=6$, 11 and 22~GeV for LT only (solid lines) and with HT corrections (bands). The lower panels show the ratios of $\Apvd$ with HT corrections to the LT $\Apvd$. The gray shaded region spans values of $\xb$ for which $W^2 < 4~{\rm GeV}^2$ that would be excluded from the global QCD analysis.}
    \label{fig:HT}
\end{figure}

In addition to the nonzero mass effects, dynamical HT effects, such as multi-parton correlations or some rescattering effects, can give corrections to structure functions at finite kinematics.
Theoretical studies of the HT effects are less developed than those of the leading, twist-2 contributions (see Refs.~\cite{Fajfer:1984um, Castorina:1985uw, Qiu:1988dn, Signal:1996ct, Dasgupta:1996hh, Stein:1998wr, Hobbs:2008mm, Mantry:2010ki, Seng:2013fia, Motyka:2017xgk}), but some phenomenological studies of the HT corrections have previously been performed~\cite{Cocuzza:2021rfn, Cerutti:2025yji, Li:2023yda, Cocuzza:2026zoy}.
In particular, we follow recent JAM analyses in using an additive HT correction term, so that for each of the structure functions $F_i$ ($i=1,2,3$) including HTs we have~\cite{Cocuzza:2026zoy},
\begin{align}
    F_{i}^{j}(\xb,Q^2) = F_{i}^{j\,(\rm LT+TMC)}(\xb,Q^2) + \frac{H_{i}^{j}(\xb)}{Q^2}
.\end{align}
While existing experimental data at low $Q^2$ provide some constraints on $H_2^\gamma$, there is little information on $H_1^\gamma$, and presently no phenomenological information at all exists on the HT $\gamma Z$ interference effects.
For our numerical analysis we model the $\gamma Z$ HT term by a multiplicative factor $R_{\rm HT}$ relative to the electromagnetic correction, $H_2^\gamma$, scaled by the appropriate leading twist functions,
\begin{align}
\label{eq:HT-def}
    H_{i}^{\gamma Z}(\xb) &= R_{\rm HT} \, H_{2}^{\gamma}(\xb) \, \frac{F_{i}^{\gamma Z}(\xb,Q^2)}{F_2^\gamma(\xb,Q^2)}
,\end{align}
and vary $R_{\rm HT}$ between $\frac12$ and 2.
The behavior of the HT corrections for the deuterium parity-violating asymmetry can be seen in \fref{HT}, becoming large as $\xb \to 1$ for fixed $Q^2$.
One can also observe the advantageous effect, with respect to these HTs, of using a higher energy beam.
For a given energy and fixed $Q^2$ one has a larger range of accessible kinematics where the HT contributions are small, and with a higher beam energy, larger $Q^2$ becomes accessible and suppresses the contribution of the HTs.

\section{Global analysis methodology}    %
\label{sec:global-analysis-methodology}  %

In this section we review the fitting methodology used in this analysis, based on the Bayesian MC global analysis framework developed in recent JAM analyses~\cite{Anderson:2024evk, Zhou:2022wzm, Cocuzza:2021cbi, Cocuzza:2026zoy}.
The PDFs are modeled at the input scale $\mu^2 = m_c^2$ using a standard functional form,
\begin{align}
    f(x;\vb*{a}) = \frac{\mathcal{N}}{\mathcal{M}}\, x^{a} (1 - x)^{b} ( 1 + c \sqrt{x} + d x )
,\end{align}
where
    $\vb*{a} = \{ \mathcal{N},a,b,c,d \}$ 
denotes the set of fit parameters, and the normalization factor 
$\mathcal{M}~=~{\rm B}[a+2,b+1] 
             + c\, {\rm B}[a+\tfrac{5}{2},b+1] 
             + d\, {\rm B}[a+3,b+1]$
is chosen to maximally decorrelate the shape and normalization parameters, with $\rm B$ the Euler beta function.
The PDF parameters are sampled via the replica method, in which an ensemble of replica experimental datasets $\widetilde{\mathcal{D}}^{(k)}$ is generated, where $k$ is the replica index, by resampling the data within their given uncertainties.
The likelihood of measuring a dataset $\mathcal{D}$, given a parameter set $\vb*{a}$ specifying the underlying theory, is given by 
    $\mathcal{L}(\vb*{a},\mathcal{D}^{(k)}) 
    = {\rm exp}\big(-\tfrac12 \chi^2 \big)$,
where
\begin{align}
    \chi^2(\vb*{a},\mathcal{D}) = \sum_{e,i} \Bigg( \frac{ d_{e,i} - \sum_k r_{e,k} \beta_{e,i}^k - T_{e,i}(\vb*{a}) / N_e }{\alpha_{e,i}} \Bigg)^2 + \sum_{e,k} r_{e,k}^2 + \sum_{e} \Big( \frac{1 - N_e}{\delta N_e} \Big)^2 \,
.\end{align}
Here, $d_{e,i}$, $\alpha_{e,i}$, and $\beta_{e,i}$ are the central value, uncorrelated uncertainty, and correlated uncertainty, respectively, associated with data point $i$ from experimental dataset $e$, and $T_{e,i}(\vb*{a})$ is the theoretical value corresponding to this data point given parameters $\vb*{a}$.
Additionally, $N_e$ and $r_{e,k}$ are nuisance parameters allowing multiplicative and additive shifts to the theory within experimental uncertainties.
A PDF replica is defined by parameters 
\begin{align}
    \vb*{a}^{(k)} = \underset{ \vb*{a} }{\rm argmin} \, \chi^2(\vb*{a},\widetilde{\mathcal{D}}^{(k)})
,\end{align}
satisfying physical constraints, such as sum rules, and upper and lower bounds.

To initialize each fit for this analysis, we set the PDF parameters to one of the replicas from a recent JAM global analysis~\cite{Cocuzza:2026zoy} and allow only the unpolarized PDF parameters to vary.
We also resample the experimental data within their given uncertainties prior to each fit.
From the replica ensemble, estimates of the mean and variance of an observable $\mathcal{O}$ are computed via
\begin{align}
{\rm E}[\mathcal{O}] &= \frac{1}{n} \sum_k \mathcal{O}(\vb*{a}^{(k)}), 
\\
{\rm V}[\mathcal{O}] &= \frac{1}{n} \sum_k \Big( \mathcal{O}(\vb*{a}^{(k)}) - {\rm E}[\mathcal{O}] \Big)^2
,\end{align}
while the correlation between observables $\mathcal{O}_1$ and $\mathcal{O}_2$ is given by
\begin{align}
\rho[\mathcal{O}_1,\mathcal{O}_2] 
&= \frac{1}{\sqrt{ {\rm V}[\mathcal{O}_1] {\rm V}[\mathcal{O}_2] }} 
\bigg[ \frac{1}{n} \sum_k 
\Big( \mathcal{O}_1(\vb*{a}^{(k)}) - {\rm E}[\mathcal{O}_1] \Big) 
\Big( \mathcal{O}_2(\vb*{a}^{(k)}) - {\rm E}[\mathcal{O}_2] \Big) 
\bigg]
.\end{align}
\\

\section{Phenomenology}   %
\label{sec:phenomenology} %

In this section we discuss parity-violating asymmetry measurements at Jefferson Lab with the existing 11~GeV and possible future 22~GeV beams using the SoLID detector, outlining the projected kinematic coverage at these energies and the simulation of $\Apv$ pseudodata and uncertainty estimates.
We perform a new global analysis within the JAM framework using these pseudodata, and discuss their impact on the $s^+$ PDF and $d/u$ PDF ratio, as well as on the weak mixing angle, $\ssw$.

\subsection{Simulation of pseudodata} %
\label{ssec:generating-pseudodata}    %

The simulated asymmetry $\Apv$ for proton and deuterium targets is shown in \fref{kinematics} for projected kinematics for the SoLID detector at the 11~GeV electron beam energy at Jefferson Lab, as well as at kinematics for a 22~GeV electron energy.
To avoid the nucleon resonance region, we make a cut on the data of $W^2 > 4~{\rm GeV}^2$.
The pseudodata are given at the bin centers $(\langle x_B \rangle_{\rm bin}, \langle Q^2 \rangle_{\rm bin})$, where the bin average of an arbitrary function $f(\xb,Q^2)$ is defined by,
\begin{align}
\label{eq:bin-centers}
    \langle f \rangle_{\rm bin} = \int_{\rm bin} \dd{\xb} \dd{Q^2} \, f(\xb,Q^2) \, a(\xb,Q^2) \frac{\diff \sigma_{UU}}{\diff \xb \, \diff Q^2} \Bigg/ \int_{\rm bin} \dd{\xb} \dd{Q^2} a(\xb,Q^2) \frac{\diff \sigma_{UU}}{\diff \xb \, \diff Q^2}
,\end{align}
where $a(\xb,Q^2)$ is the detector acceptance, simulated on the dense kinematics shown in the left panel of \fref{kinematics}.

\begin{figure}[t]
    \centering
    \includegraphics[width=\textwidth]{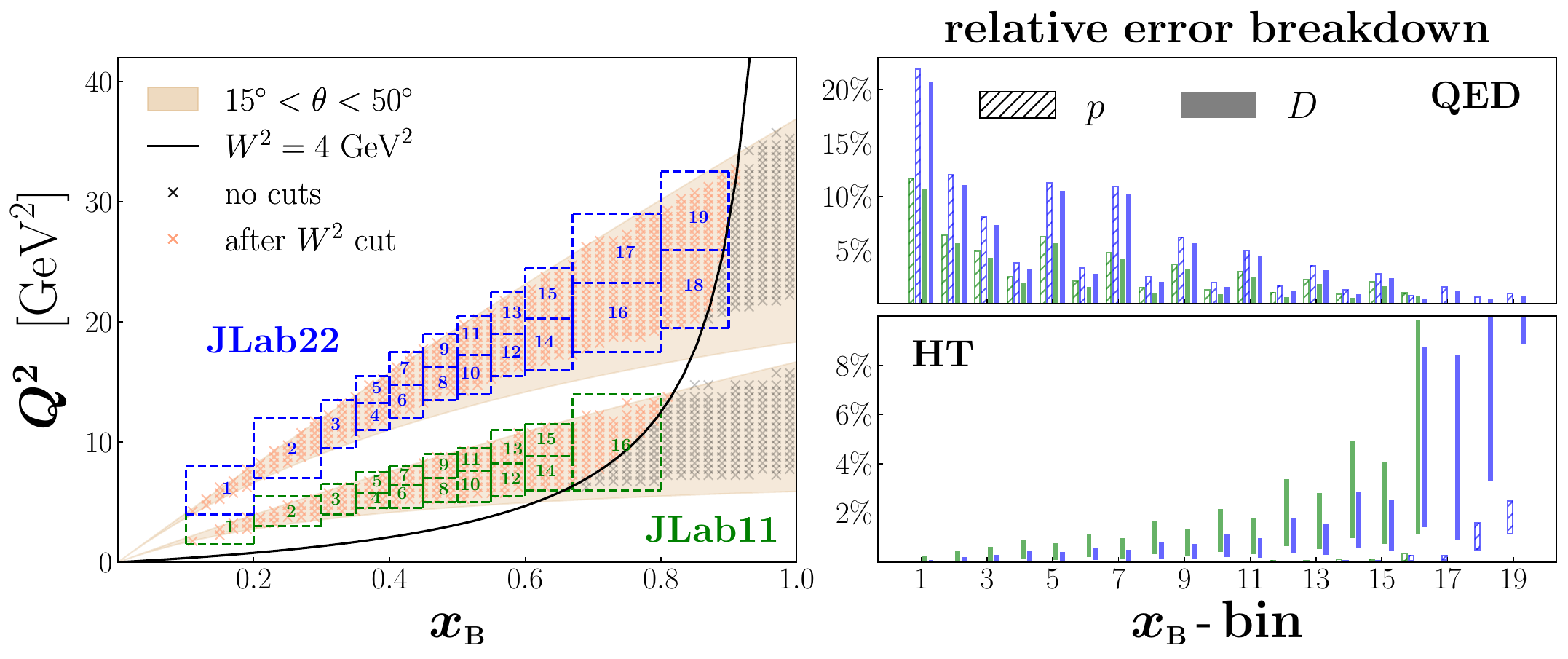}
    \caption{[Left] Projected experimental kinematics for the SoLID detector at 11~GeV and 22~GeV electron energies. The light brown region approximates the polar angle acceptance, within which the ``$\times$''s represent the $(\xb,Q^2)$ points where the detector acceptance is simulated. The light red markers represent simulated kinematics that survive the $W^2 > 4~{\rm GeV}^2$ cut and are subsequently placed into the 11~GeV (22~GeV) bins indicated by blue (green) dashed lines. The $\Apv$ pseudodata are defined at the bin centers indicated by the position of the numbers and determined via \eref{bin-centers}. 
    [Right] Breakdown of the QED (upper panel) and HT (lower panel) theoretical relative uncertainties within each kinematic bin for the proton (hatched) and deuterium (solid) $\Apv$ pseudodata shown in the left panel. For the HT effects, the bars span the range of the uncertainties from $R_{\rm HT}=\frac12$~to~2 [see \eref{HT-def}]. Note that the green and blue bars correspond to the 11~GeV and 22 GeV kinematics of the left panel, respectively.}
    \label{fig:kinematics}
\end{figure}

The central value for $\Apv^{p}$ and $\Apvd$ in a bin is computed with a single PDF replica which most closely reproduces the mean values across all bins.
Defining the central values of the pseudodata using the single PDF replica, instead of expected values computed across the ensemble of replicas, avoids any biases arising through nonlinear effects from averaging the parity-violating asymmetry.
Statistical uncertainties are assigned via
\begin{align}
\label{eq:statistical-uncertainty}
    \delta^{\rm stat} \Apv = \frac{1}{P \sqrt{N}}
,\end{align}
where $P$ is the beam polarization, estimated to be 85\%~\cite{JeffersonLabSoLID:2022iod}.
The number of events in a bin
\begin{align}
    N = \mathcal{L} \sigma_{\rm bin} = \mathcal{L} \int_{\rm bin} \diff \xb \, \diff Q^2 \, a(\xb,Q^2) \frac{\diff \sigma}{\diff \xb \, \diff Q^2}\
    \approx\ \mathcal{L} \sum_i a_i \sigma_{i}
,\end{align}
where $\mathcal{L}$ is the integrated luminosity.
For our simulations we use $\diff \mathcal{L} / \diff t$ \mbox{$= 4.85 \times 10^{38}~{\rm cm^{-2}~s^{-1}}$} (achievable with a $45~\mu{\rm A}$ beam on a $40~{\rm cm}$ long target) and a run-time of 50 days for each target.
On top of the statistical uncertainty, a baseline experimental correlated systematic uncertainty of 0.5\% is assigned to each data point~\cite{JeffersonLabSoLID:2022iod}.
Finally, theoretical uncorrelated systematic uncertainties, from QED and HT effects, are included as
\begin{subequations}
\begin{align}
\label{eq:QED-HT-uncertainty}
    \delta^{\rm QED} \Apv &= \Big| \Apv^{(\rm RC)} - \Apv^{(\rm no \, RC)} \Big|,
    \\
    \delta^{\rm HT} \Apv  &= \Big| \Apv^{(\rm LT + HT)} - A_{\rm PV}^{(\rm LT)} \Big|
.\end{align}
\end{subequations}

In a comprehensive global analysis with real experimental data, one should simultaneously fit the LDFs and LFFs, which also contain nonperturbative hadronic contributions within the QED+QCD hybrid factorization framework, together with the $\gamma Z$ interference HT functions, in order to extract these functions consistently alongside the PDFs and $\ssw$~\cite{Cammarota:2025jyr}.
In the present work, however, the primary goal is to assess the potential impact of the $\Apv$ measurements, rather than to perform a full-scale global fit.
It therefore suffices to estimate the magnitude of the QED and HT contributions and treat them as systematic background uncertainties, modeling how they dilute the sensitivity of $\Apv$ data to the underlying electroweak and PDF parameters.

In the right column of \fref{kinematics}, we display these uncertainties for each corresponding bin in the left-hand panel of the figure.
The QED uncertainties are shown in the upper-right panel.
Consistent with \fref{Apv-RC}, we observe that these effects become small as the bin index increases, corresponding generally to larger $\xb$ and $Q^2$ values.
However, comparing the projected kinematics for the 11~GeV and 22~GeV pseudodatasets, we find that the QED uncertainties are typically larger for the latter.
This is attributable to the fact that the inelasticity must increase to keep the electron scattering angle fixed at larger beam energies, so that in turn the projected 22~GeV kinematics lie in non-optimal regions with respect to the QED effects.
It may be possible, however, to adjust the experimental configuration to access kinematics where these uncertainties are reduced without affecting the measured range of $\xb$.

The HT uncertainties for the pseudodata are shown in the lower-right panel of \fref{kinematics}.
Note that the bars represent the range of uncertainties corresponding to the variation of the $R_{\rm HT}$ parameter defined in \eref{HT-def}, where the lower and upper ends of the bar corresponds to the minimum and maximum uncertainties from this variation.
As expected, the uncertainties increase as the bin index increases, and the 22~GeV uncertainties tend to be smaller than those of the 11~GeV pseudodata, as one would expect.
Furthermore, the HT uncertainties tend to be smaller for the proton pseudodata, which should improve constraints on the $d/u$ PDF ratio at large parton momentum fractions.
Finally, with respect to the QED uncertainties, modifying the kinematics for the 22~GeV pseudodata as described above should not significantly enhance the HT uncertainties. \\

\subsection{Impact on PDFs and the Weinberg angle} %
\label{ssec:impact-on-strange-pdfs-weinberg-angle}         %

To fit the generated pseudodata, we use the Bayesian Monte Carlo replica methodology outlined in \sref{global-analysis-methodology} and in Refs.~\cite{Sato:2016tuz,Zhou:2022wzm}.
As a baseline we utilize the unpolarized PDFs from the recent JAM26 analysis~\cite{Cocuzza:2026zoy}, which also included HT parameters when fitting lower energy data.
To see the effects of the different sources of uncertainty, we fit pseudodatasets first containing only the statistical uncertainties from \eref{statistical-uncertainty} and second with the experimental and theoretical systematic uncertainties added.
These fits are performed with the 11~GeV dataset and combined 11 and 22~GeV datasets, which allows us to better diagnose whether the observed impacts are from statistics or systematics.

The impact of the $\Apv$ pseudodata on the uncertainties of the strange quark PDFs and the $d/u$ ratio is illustrated in \fref{s+-impact}.
Clearly the $\Apv$ asymmetry with only a nominal statistical uncertainty has a strong constraining power on the strange PDFs and $d/u$ ratio, particularly on the former, with a reduction up to $\approx 30\%$ from the 11~GeV pseudodatasets and 50\% from the 22~GeV pseudodatasets.
Interestingly, although the 11~GeV kinematics may be expected to yield better statistical precision due to the cross section being enhanced for a fixed beam energy at lower $Q^2$, since the differential cross section scales as $s/Q^4$, the squared center of mass energy compensates the suppression at larger $Q^2$.
In addition, the fiducial volume of the 22~GeV phase space tends to be bigger, allowing larger cross sections in each bin for a fixed run-time.

Upon the addition of the systematic uncertainties, the impact of the $\Apv$ pseudodata generally weakens.
At smaller momentum fractions, the uncertainties arising from QED effects (see \fref{kinematics}) wash out much of the signal for both 11~GeV and 22~GeV kinematics.
On the other hand, at larger momentum fractions the fits with the 22~GeV pseudodata show distinct improvement relative to those with only the 11~GeV pseudodata, stemming from the accessibility of kinematics where the HT contributions to $\Apv$ are suppressed.

\begin{figure}[t]
    \centering
    \includegraphics[width=0.63\textwidth]{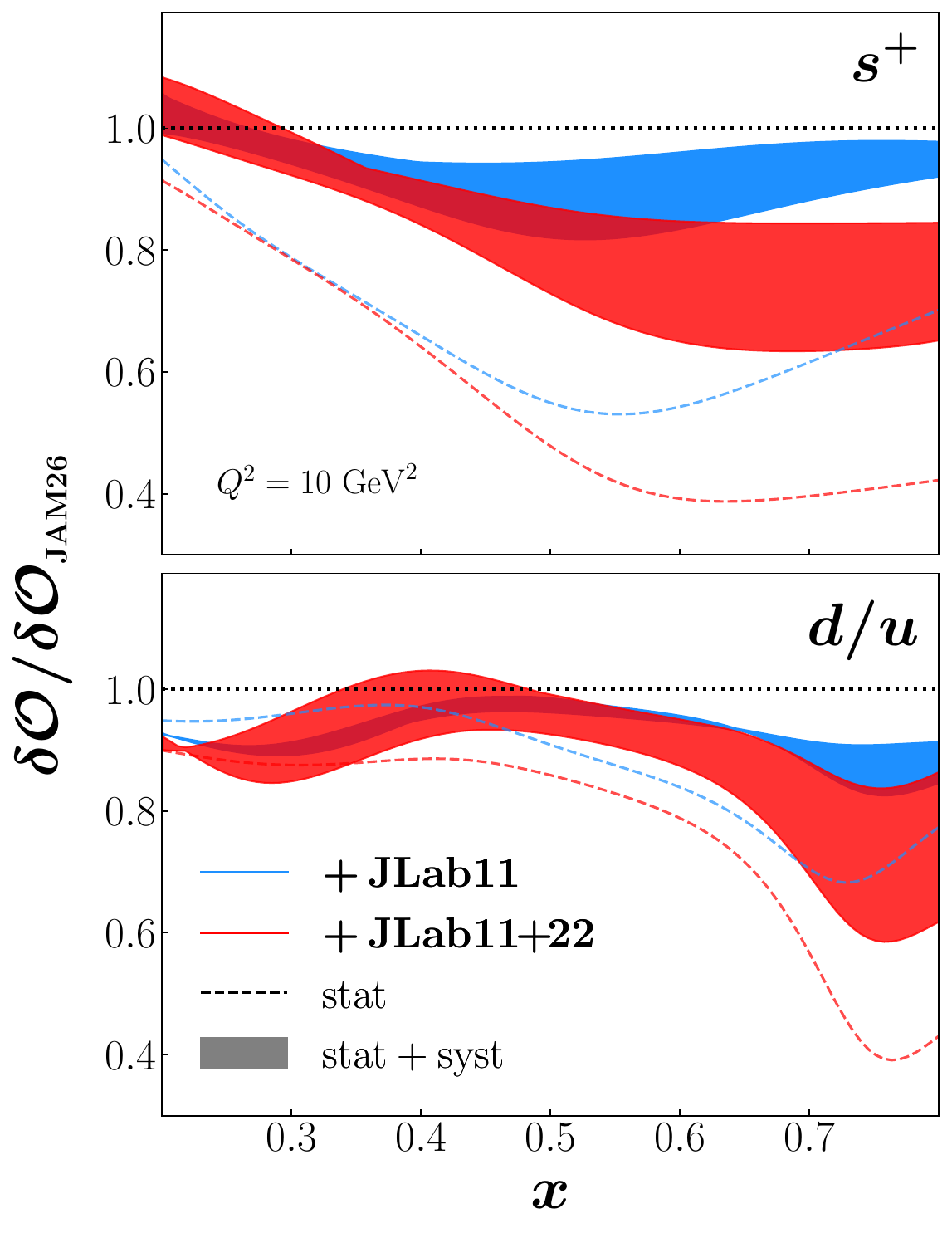}
    \caption{
    Uncertainty on the $s^{+}$ PDF and the $d/u$ PDF ratio relative to the baseline JAM26~\cite{Cocuzza:2026zoy} PDF uncertainty at $Q^2=10$~GeV$^2$ after fits with combined proton and deuterium $\Apv$ datasets with statistical uncertainties only (dashed lines) and with full statistical and systematic uncertainties (bands). The widths of the bands correspond to variations of $R_{\rm HT}$ between $\frac12$ and 2 [see~\eref{HT-def}].
    }
    \label{fig:s+-impact}
\end{figure}

As expected from the LO form of the parity-violating asymmetry, the constraints on the strange PDFs come primarily from the deuterium pseudodata.
There is some joint constraint from the proton data, which can be identified through the following combination of proton and deuteron structure functions~\cite{Accardi:2023chb},
\begin{align}
\label{eq:cal-F} 
    5 F_{2}^{\gamma Z\, p} - 4 F_{2}^{\gamma\, D} 
    \approx \frac23 xs^{+} - \frac23 xc^{+} 
    + \frac{20 \Delta \ssw}{9} \big( 4 xu^{+} + xd^+ + xs^+ + 4 xc^+ \big)
,\end{align}
where $\Delta \ssw = \frac14 - \ssw$.
Note, however, that the first 2 terms $\sim xs^+ - xc^+$ (and we can assume $s^+ \gg c^+$ at large $x$) appear alone only under the assumption of maximal mixing in the weak sector.
In practice, the terms proportional to $\Delta \ssw$ can lead to significant deviations from a purely strange signal.
From our analysis, we find a sizable constraint on this observable when only statistical uncertainties are present, but the presence of systematic uncertainties renders the pseudodata ineffective toward constraining this observable.
This suggests that the strange signal from proton pseudodata, which is relatively weak compared to that of deuterium, is contaminated by the presence of the systematic uncertainties.
On the other hand, the proton pseudodata are essential for the constraint on~$d/u$.

In this work we also fit the weak mixing angle, fixing the unpolarized PDFs.
For this, we use the SM running of the coupling, treating the value of $\ssw$ at the $Z$-boson mass as a free parameter.
Traditionally, this parameter is determined from high-energy collider data, which yield very precise determinations of $\ssw$.
Our goal here, however, is to understand the ability of future low-$Q^2$ data from SoLID to discriminate between measurements of $\ssw$ at separate ends of the energy spectrum, which are currently in tension.

\begin{figure}[t]
    \centering
    \includegraphics[width=0.75\textwidth]{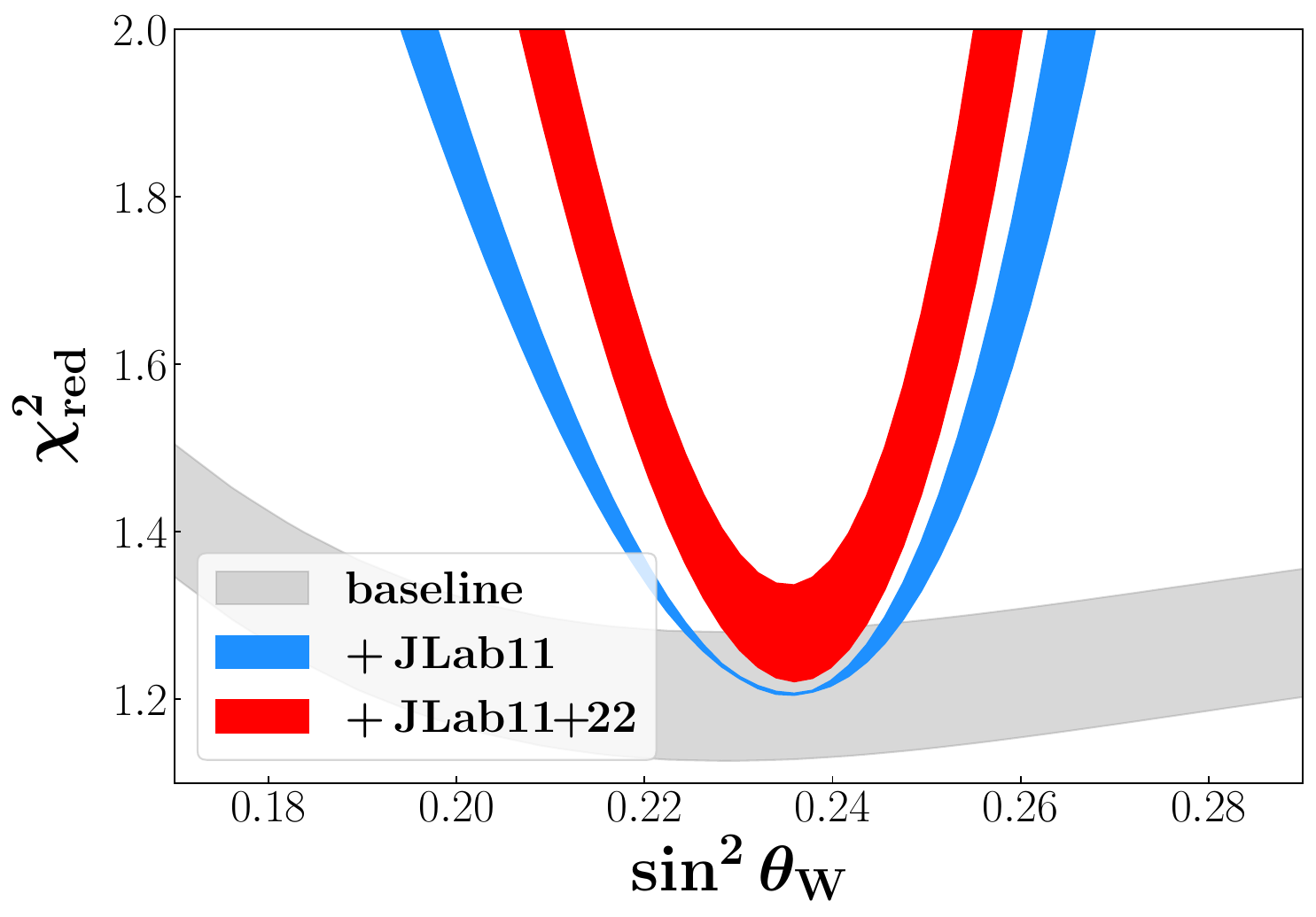}
    \caption{
    Reduced-$\chi^2$ profile as a function of $\ssw$, evaluated at $Q^2 = 4$~GeV$^2$ using the SM running. The baseline results using the JAM26 PDFs~\cite{Cocuzza:2026zoy}, supplemented with SLAC~\cite{Prescott:1978tm, Prescott:1979dh} and 6~GeV Jefferson Lab $\Apv$ data~\cite{Wang:2014guo} (gray), are compared with those including projected Jefferson Lab~11~GeV data (blue) and the combined projected 11~GeV and 22~GeV data (red).
    }
    \label{fig:rchi2-likelihood-s2w-running}
\end{figure}

To this end, we use legacy $\Apv$ data from SLAC~\cite{Prescott:1978tm, Prescott:1979dh} and Jefferson Lab at 6~GeV~\cite{PVDIS:2014cmd} to establish a baseline prior distribution for $\ssw$ at $Q=M_Z$.
We then add the Jefferson Lab $\Apv$ pseudodata and perform Lagrange multiplier scans of the weak mixing angle, the results of which are shown in \fref{rchi2-likelihood-s2w-running} at $Q^2 = 4~{\rm GeV}^2$.
The reduced-$\chi^2$ profile makes quantitative our observations from \ssref{sensitivity-on-strange-PDFs-ssw} about the sensitivity of $\Apv$ to $\ssw$.
Since the value of the parity-violating asymmetry is essentially directly governed by the weak mixing angle at the relevant scale, the constraining power on $\ssw$ from $\Apv$ is limited purely from the size of the uncertainties.
Statistical uncertainties dominate the SLAC $\Apv$ measurements, and the 6~GeV Jefferson Lab $\Apv$ data, albeit precise, consist of only a few data points.
Consequently, the constraints from existing data are limited, and our results here signal a strong impetus for future measurements of~$\Apv$.

In the preceding analyses, we assessed the impact of the $\Apv$ data on $\ssw$ and $s^{+}$ individually, either by fixing $\ssw$ at the $Z$-boson mass while fitting the PDFs, or by fitting $\ssw$ while keeping the PDFs fixed.
From the theoretical expressions, it is clear that the PDFs and $\ssw$ are directly coupled, although it is not necessarily clear \textit{a priori} to what extent they are statistically correlated.
In order to assess fully the combined impact of the $\Apv$ data on $s^+$ and $\ssw$, we next allow the PDF and weak mixing angle parameters to vary simultaneously.

\begin{figure}[t]
    \centering
    \includegraphics[width=0.95\textwidth]{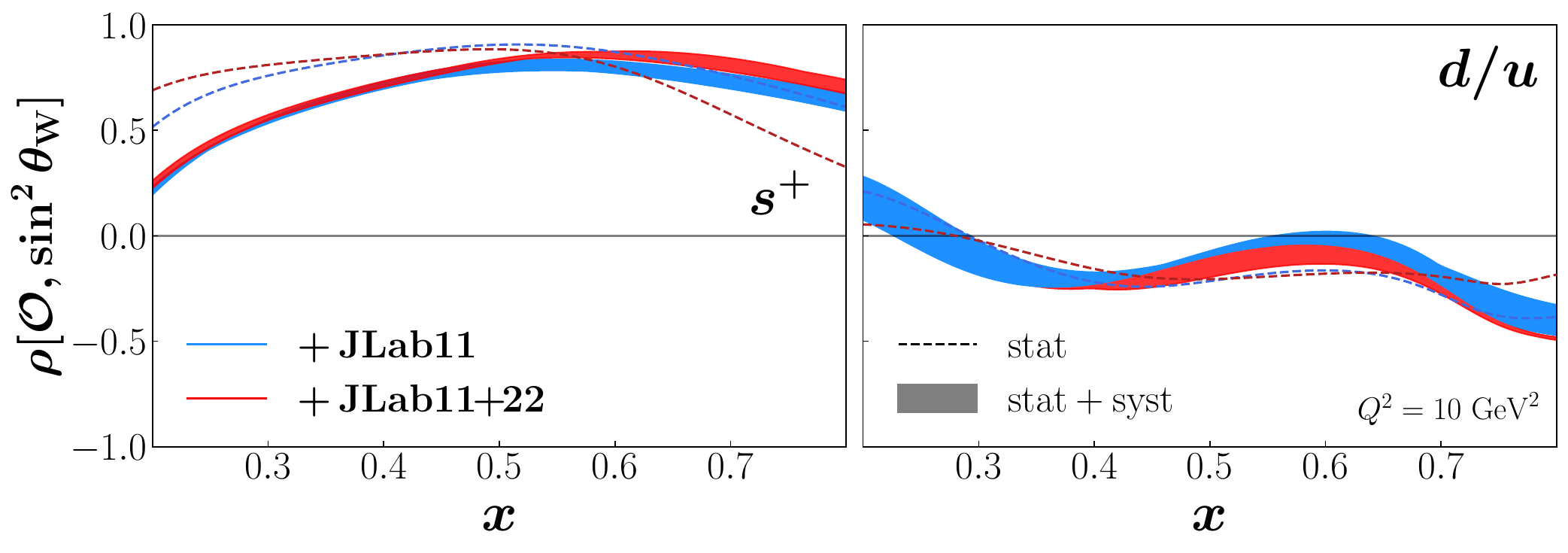}
    \caption{
    Correlation $\rho$ between the weak mixing angle, $\ssw$, and the strange quark PDF $s^{+}$ (left) and the $d/u$ quark PDF ratio (right) using Jefferson Lab 11~GeV pseudodata (blue) and the combined 11~GeV and 22~GeV pseudodata (red). The full results (bands) are compared with those using statistical uncertainties only (dashed lines).
    }
    \label{fig:PDFs-sin2w-corr}
\end{figure}
\begin{figure}[h]
    \centering
    \includegraphics[width=0.95\textwidth]{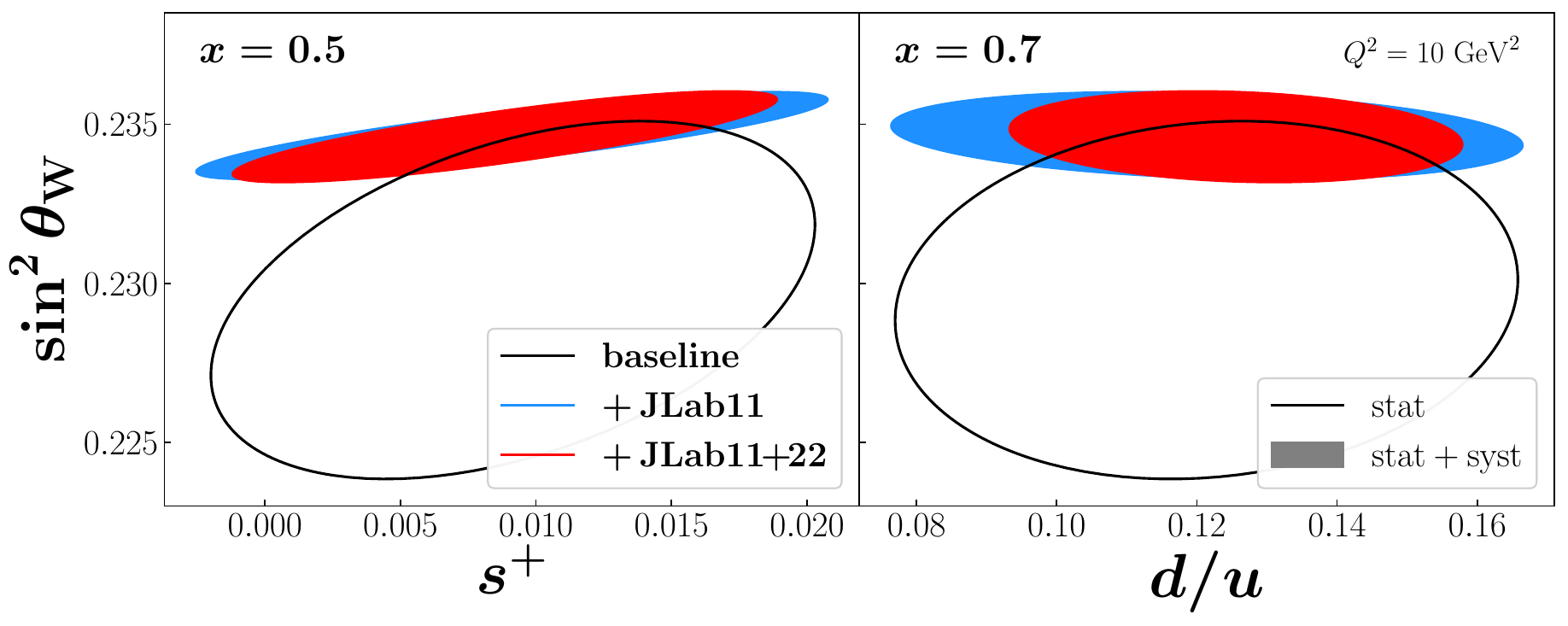}
    \caption{
    Joint $68\%$ confidence regions for $\ssw$ with the $s^+$ PDF at $x=0.5$ (left) and with the $d/u$ PDF ratio at $x=0.7$ (right) at $Q^2 = 4$~GeV$^2$ from a combined fit of PDFs and $\ssw$ using Jefferson Lab 11~GeV pseudodata (blue) and the combined 11~GeV and 22~GeV pseudodata (red), compared with the baseline (black), which is a fit including data from the JAM26 analysis and $A_{\rm PV}$ data from the 6 GeV JLab~\cite{Wang:2014guo} and SLAC~\cite{Prescott:1978tm,Prescott:1979dh}.
    }
    \label{fig:PDFs-sin2w-impact}
\end{figure}

As could be expected from the approximate LO expressions for $A_{\rm PV}$ in \eref{Apv-D-LO}, we find that a sizable, positive correlation exists between the $s^{+}$ PDF and $\ssw$ across a wide range of momentum fractions, illustrated in \fref{PDFs-sin2w-corr}.
This may limit the joint constraint of the two quantities to some degree, as seen in \fref{PDFs-sin2w-impact}.
It is clear, in any case, that $\Apv$ input within a global analysis provides a clean signal for the value of $\ssw$, capable of fixing the central value of the former much more precisely than existing low-$Q^2$ PVDIS data.

On the other hand, the correlation between $d/u$ and $\ssw$ appears milder across most smaller values of $x$, but can become approximately $-50\%$ at very large $x$, $x \approx 0.8$, attributable to the multiplicative contribution of $d/u$ and $\ssw$ to $\Apv^{p}$.
The proton $\Apv$ data therefore permit a stronger constraint of the $d/u$ PDF ratio.
The addition of 22~GeV data also yields significantly smaller uncertainties than just the 11~GeV data.
This trend is apparent for both $s^+$ and $d/u$, signaling that high statistics and broad kinematic coverage will be necessary to constrain the PDFs alongside the electroweak parameters.

\begin{figure}[t]
    \centering
    \includegraphics[width=0.85\linewidth]{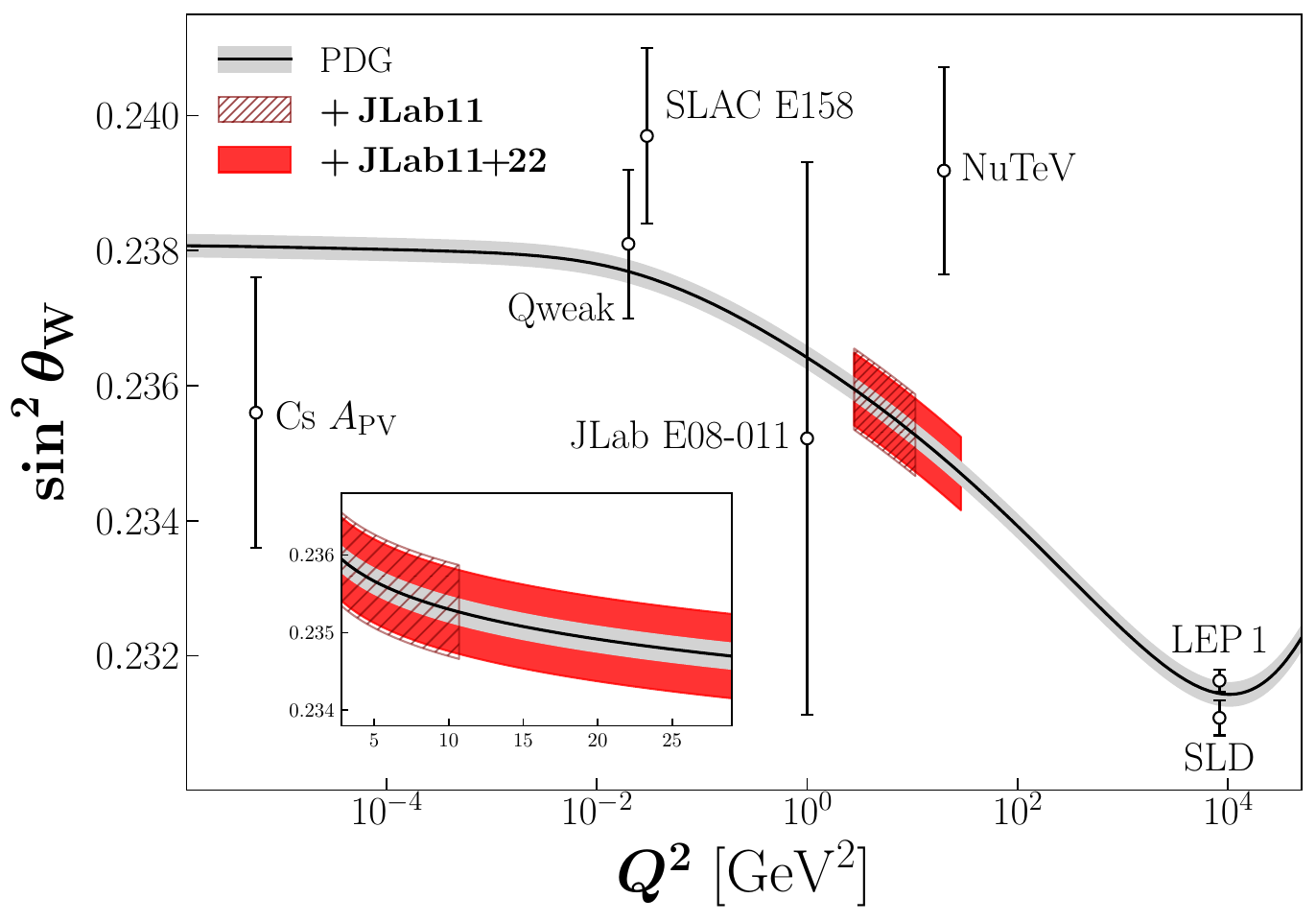}
    \caption{Running of $\ssw$ (in the $\overline{\rm MS}$ scheme) with energy scale $Q^2$ compared with selected experimental measurements. The value at the $Z$ pole is provided by Ref.~\cite{ParticleDataGroup:2020ssz} through an average of the $\rm LEP~1$ and $\rm SLD$ measurements. The red bands show uncertainties for $\ssw$ from simultaneous fits of $\ssw$ and unpolarized PDFs to simulated SoLID pseudodata across the range of $Q^2$ permitted by the pseudodata.}
    \label{fig:s2w-RGE-measurements}
\end{figure}

Finally, in \fref{s2w-RGE-measurements}, we show a projection of the constraints on the running of $\ssw$ with energy across the range of $Q^2$ provided by the 11~GeV and 22~GeV Jefferson Lab datasets.
These expectations are compared with the existing experimental results and the running predicted in the SM from the combined LEP~1 and SLD value for $\ssw$ at the $Z$-pole~\cite{ParticleDataGroup:2020ssz}.
(Interestingly, LEP~1 and SLD measurements themselves appear in tension at the level of $3\sigma$.)
While the extracted central value for $\ssw$ in this study is biased by our simulation of $\Apv$ pseudodata using the PDG SM results, the small extracted uncertainties illustrate the potential discriminating power afforded by $\Apv$ data.
That is, future PVDIS measurements at Jefferson Lab can provide the most precise test of the SM in the modern era of high energy and nuclear physics.

\section{Conclusion}   %
\label{sec:conclusion} %

We have presented a comprehensive analysis of the parity-violating asymmetry in the DIS of polarized electrons from unpolarized proton and deuteron targets, illustrating the unique sensitivity of future Jefferson Lab data to the strange quark PDF at small and intermediate momentum fractions, as well as the $d/u$ ratio in the high-$x$ region, and the weak mixing angle $\ssw$ at low $Q^2$.
Using the JAM Bayesian global QCD analysis framework, we investigated the impact of incorporating $\Apv$ pseudodata from simulated kinematics of the 11~GeV SoLID detector, and those with an upgraded 22~GeV beam at Jefferson Lab.
Importantly, background effects from radiative corrections using the QED+QCD hybrid factorization~\cite{Liu:2021jfp} and models of $\gamma Z$ HT have been studied, and can be sizable at certain kinematics, indicating the importance of future studies dedicated to their phenomenological impact.
These fits demonstrate that the PVDIS asymmetry possesses a considerable constraining power on the size of the nucleon strangeness and the $d/u$ PDF ratio.
Our analysis also reaffirms the significant constraining power that $\Apv$ data can have on the value of the weak mixing angle, providing a fundamental test of the Standard Model at low~$Q^2$.

Looking forward, there are several natural avenues to extend this work and address the outstanding questions discussed here regarding nucleon structure and fundamental physics.
First, in this study we modeled the deuteron PDFs as linear combinations of the proton and neutron PDFs under the assumption of exact isospin symmetry.
While the nuclear effects are expected to partially cancel between the numerator and denominator in the asymmetry, a systematic treatment of these corrections will be essential for precision extractions of the $d/u$ ratio.
Moreover, charge symmetry violation, arising from quark mass differences and electromagnetic effects, can produce small but non-negligible deviations at the asymmetry level~\cite{Hobbs:2008mm, Wang:2024mll}, potentially impacting the extraction of $d/u$ and its interpretation in a global analysis.
Finally, calculations of the NLO hard parts in the hybrid QED+QCD factorization were recently completed~\cite{Cammarota:2025jyr}, increasing the perturbative accuracy of the radiative corrections and enabling future global analyses incorporating LDFs and LFFs.

Because of the kinematic suppression of $F_3^{\gamma Z}$ in the parity-violating asymmetry, measurements exploiting the charge dependence of the unpolarized cross section will be essential for constraining the unpolarized strange--antistrange asymmetry and elucidating the role of sea quarks in the nucleon, particularly relating to nonperturbative dynamics~\cite{Accardi:2020swt}.
Semi-inclusive scattering data with identified pions and kaons will enable future global analyses to leverage enhanced flavor separation and better isolate specific PDF flavor combinations.
Given the elusive nature of the partonic sea in experimental settings, additional input from lattice QCD calculations directly sensitive to the strange-quark PDFs may be valuable~\cite{Yang:2018nqn, Alexandrou:2020sml}, and recent studies have demonstrated the utility of lattice calculations as theoretical supplements to experimental data in global analyses~\cite{Barry:2025wjx, Good:2025nny, Gamberg:2022kdb, JeffersonLabAngularMomentumJAM:2022aix, Bringewatt:2020ixn, Lin:2017stx}.
Finally, neutrino DIS can provide complementary sensitivity to sea-quark dynamics through charged current flavor mixing, offering additional opportunities to test the SM in global analyses via precision determinations of the Cabibbo–Kobayashi–Maskawa matrix elements and electroweak couplings.
However, since neutrino scattering almost universally takes place on nuclear targets, additional care must be taken to account for nuclear effects in charged current processes, which are generally not as well understood as those in electromagnetic interactions.

In the polarized PDF sector, recent global QCD analyses indicate a lack of data-driven constraint on the helicity-dependent strange quark PDF $\Delta s^{+}$, including its overall sign and magnitude~\cite{Zhou:2022wzm, Hunt-Smith:2024khs, Cocuzza:2025qvf}.
Although more challenging experimentally than the PV asymmetry measurements discussed here, scattering unpolarized electrons from longitudinally polarized protons or deuterons can also yield sensitivity to $\Delta s^{+}$ in the polarized PVDIS asymmetry, $\Delta \Apv$, which is defined in analogy to $\Apv$~\cite{Anselmino:1993tc, Hobbs:2008mm, Boughezal:2022pmb}.
Crucially, this sensitivity is linear in $\Delta s^{+}$ since only the numerator depends on the polarized structure functions and PDFs.
An impact study for the Electron-Ion Collider, where hadronic beams can be more easily polarized, has been conducted, although not yet within the context of a global analysis~\cite{Boughezal:2022pmb}.
Given that Jefferson Lab provides the highest luminosity of any active or planned facility, prospects for a polarized target and impact studies of $\Delta \Apv$ would be of considerable interest.

\begin{acknowledgements}

We thank Jianwei Qiu, Trey Anderson and Christina Cocuzza for helpful discussions and communications.
This material is based upon work supported by the U.S. Department of Energy, Office of Science, Office of Nuclear Physics under Contract No. 89243126CSC000213, and the U.S. Department of Energy, Office of Science, Office of Workforce Development for Teachers and Scientists, Office of Science Graduate Student Research (SCGSR) program. 
The SCGSR program is administered by the Oak Ridge Institute for Science and Education (ORISE) for the DOE, which is managed by ORAU under contract number DESC0014664.
The work of R.~M.~W. was partially supported by a Jefferson Science Associates (JSA) Graduate Fellowship.
The work of T.~L. was supported by the National Key R\&D Program of China No. 2024YFA1611004 and by the National Natural Science Foundation of China No. 12321005. 
The work of N.~S. was supported by the DOE, Office of Science, Office of Nuclear Physics in the Early Career Program.

\end{acknowledgements}

\bibliography{bibliography.bib}

\end{document}